\documentclass[11pt,a4paper]{article}
\usepackage{jheppub}
\usepackage{epsfig}
\newcommand{\be}{\begin{equation}}
\newcommand{\ee}{\end{equation}}
\newcommand{\bea}{\begin{eqnarray}}
\newcommand{\eea}{\end{eqnarray}}

\def\r2{{\tilde r}}
\def\Li{{\rm Li}}

\title{Next-to-leading order contributions to the pole mass of gluino in minimal gauge mediation}

\author[a]{Jae~Yong~Lee,}
\author[b]{and Yeo~Woong~Yoon}

\affiliation[a]{
Department of Physics, Korea University,\\
Seoul 136-701, Korea}
\affiliation[b]{School of Physics, KIAS,\\
Seoul 130-722, Korea}

\emailAdd{littlehiggs@korea.ac.kr}
\emailAdd{ywyoon@kias.re.kr}
\abstract{We compute the pole mass of the gluino in the minimal gauge mediation to two-loop order.
The pole mass of the gluino begins to arise at one-loop order and
the two-loop order correction shifts the leading order pole mass by 20\% or even more.
This shift is much larger than the expected accuracy of the mass determination at the LHC,
and should be reckoned with for precision studies on the SUSY breaking parameters.}
\arxivnumber{1112.3904}

\begin{document}
\maketitle

\section{Introduction}
Supersymmetry, an elegant extension of spacetime symmetry,
is the leading candidate for the new physics unfolded in the Large Hadron Collider (LHC) at CERN.
The two detectors ATLAS and CMS of the LHC have been collecting data at a much faster than expected
and their recent data significantly extend the exclusion limits for supersymmetric particles.
But the latest such data have so far been interpreted by the experiment in only two different
supersymmetry breaking models: the constrained minimal supersymmetric standard model (CMSSM)
and a simplified model with only squarks and gluions and massless neutralinos.

Other supersymmetry breaking models should be extensively analyzed in the era of the LHC.
One of those to be analyzed is gauge mediated supersymmetry breaking (GMSB)
model~\cite{Dine:1981rt,Dimopoulos:1981au,Dine:1981gu,Nappi:1982hm,
AlvarezGaume:1981wy,Dimopoulos:1982gm,Dine:1993yw,Dine:1994vc,Dine:1995ag}
where messenger fields, charged under the Standard Model gauge symmetry,
mediate the breakdown of supersymmetry in the hidden sector to the MSSM sector.
The soft masses in the visible sector arise from quantum effects of the messengers
so the supersymmetry breaking scale of the visible sector is much lowered
than the grand unified theory (GUT) scale compared with the gravity mediated supersymmetry breaking scenario.

As the LHC continues to collect experimental data, the precise studies on the physical parameters
of the SUSY particles will become important.
This requires one to do quantum loop calculations on the SUSY parameters.
For instance, the gluino pole mass of the CMSSM has been considered up to two-loop order
in ref.~\cite{Martin:2005ch,Yamada:2005ua,Martin:2005eg} while the neutralino and chargino
pole masses in ref.~\cite{hep-ph/0612276,arXiv:0706.0781}.

Our interests lie in the gluino pole mass of the GMSB model at two-loop order.
Among the various different GMSB models we choose the Minimal Gauge Mediation (MGM)
where a pair of messenger fields, fundamental and antifundamental under the $SU(3)_C$ gauge symmetry,
mediate the supersymmetry breaking to the MSSM sector.
The gluino pole mass of the GMSB model at two-loop order was first discussed in ref.~\cite{Picariello:1998dy}
where the authors made a prediction that the NLO correction to the gluino pole mass is up to 10\% of the LO pole mass.
On the other hand, our prediction is, as shown later, 20\% or even more.
The difference between their and our predictions arises from how to handle the IR behavior of SUSY QCD.
We state that our treatment of the IR behavior is more consistent with perturbative theory rather than theirs.
We will rigorously discuss it in Section~\ref{sec:selfe}.

For the renormalization of the MGM lagrangian parameters the $\overline{\rm DR}$ scheme is
adopted~\cite{Siegel:1979wq,Capper:1979ns,Jack:1993ws}. It is based on regularization by dimensional
reduction along with modified minimal subtraction($\overline{\rm MS}$) scheme.
Not only the messenger fields wavefunctions but also their masses are renormalized.
The MSSM quarks and squarks contribute to the gluino pole mass at two-loop order
through the renormalization of the gluon(gluino) wavefunctions and the gauge coupling at one-loop order.
We will also take these contributions into account.

In this paper, we follow the two-component formalism to derive the self-energy functions
in ref.~\cite{Martin:2005ch,Dreiner:2008tw}, and then present the analytic results of the self-energy functions
up to two-loop order relevant to the gluino pole mass.
We also perform a numerical analysis for the NLO correction of the gluino pole mass.

\section{Self-energy functions and pole masses for two-component spinors}

We briefly review the self-energy functions for fermions in two-component notation
and then describe how to compute the loop-corrected gluino pole mass of the MGM.
All the details can be found in ref.~\cite{Dreiner:2008tw}.

\begin{figure}[t]
\centering
\includegraphics[width=6.2in]{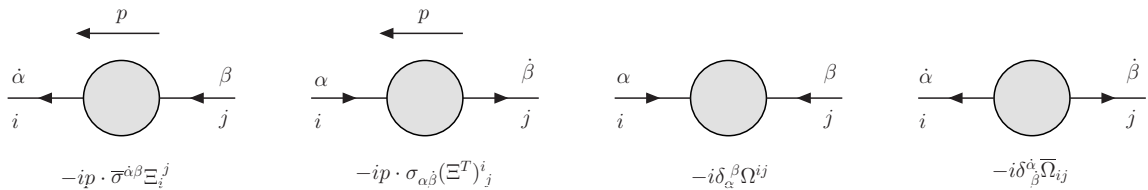}
\caption{The self-energy functions for two-component fermions.}
\label{fig:selfenergy}
\end{figure}

We consider a theory with left-handed fermion degrees of freedom $\psi_j$ with an index $j=1,2,\cdots,N$.
The self-energy functions for fermions in two-component notation are depicted in figure~\ref{fig:selfenergy},
where the shaded circles denote the sum of all one-particle irreducible(1PI), connected Feynman diagrams,
and the external legs are amputated~\footnote{At this stage, counterterm corrections are not reckoned with.}.
In figure~\ref{fig:selfenergy} the self-energy functions for two-component fermions are denoted by
$\boldsymbol{\Xi},\boldsymbol{\Xi}^T,\boldsymbol{\Omega}$, and $\overline{\boldsymbol{\Omega}}$,
and the four-momentum $p$ flows from right to left.

The pole mass is defined by the position of the complex pole in the propagator
and is a gauge-invariant and renormalization scale-invariant quantity.
The pole mass of a fermion can be found by considering its rest frame,
in which the space components of the external momentum $p^\mu$ vanish.
This reduces the spinor index dependence to a triviality. Setting $p^\mu=(\sqrt{s},0)$,
we search for the values of $s$.
In other words, the poles of the full propagator (which are in general complex)
\begin{equation}
s_{pole,j}\equiv (\boldsymbol{M}_j-i\boldsymbol{\Gamma}_j/2)^2
\end{equation}
are formally the solutions to the non-linear equation~\footnote{Here
$\boldsymbol{M}_j$ and $\boldsymbol{\Gamma}_j$
are a physical mass and width of a fermion.}
\begin{equation}
\mbox{det} [s\boldsymbol{1} -(\boldsymbol{1}-\boldsymbol{\Xi}^T)^{-1}(\boldsymbol{m}+\boldsymbol{\Omega})
(\boldsymbol{1}-\boldsymbol{\Xi})^{-1}(\overline{\boldsymbol{m}}+\overline{\boldsymbol{\Omega}})]=0,
\end{equation}
where $\boldsymbol{m}_{ij}$ is the symmetric $N\times N$ tree-level fermion mass matrix with
$\boldsymbol{m}_{ik} \overline{\boldsymbol{m}}^{kj}=m_i^2\delta^j_i$.

The gluino pole mass of the MGM is zero at tree-level and arises at one-loop order.
Moreover the gluino do not mix other fermions so the master equation
reduces to a simple equation as
\begin{equation}\label{eq:master}
s-\frac{\Omega^2}{(1-\Xi)^2}=0.
\end{equation}
The solution for Eq.~(\ref{eq:master}) is given as
\begin{equation}
\sqrt{s}=\frac{\Omega}{1-\Xi},
\end{equation}
and is perturbatively calculated as
\begin{equation}
\label{eq:polemass}
\sqrt{s}= \Omega^{(1)}+[\Omega^{(1)}\Xi^{(1)}+\Omega^{(2)}]+\cdots,
\end{equation}
where the self-energy functions are expanded in powers of $\alpha_s$:
\begin{align}
\Omega &=\Omega^{(1)}+\Omega^{(2)}+\cdots,\\
\Xi &=\Xi^{(1)}+\Xi^{(2)}+\cdots.
\end{align}

We use an iteration method to solve Eq. (\ref{eq:polemass}).
We first get the leading order (LO) pole mass by substituting the
tree level gluino mass ($s=0$) to the one-loop function $\Omega(s)^{(1)}$. Then we
substitute the LO pole mass into Eq. (\ref{eq:polemass}) to calculate the NLO pole mass.
In order to evaluate the gluino pole mass up to two-loop order
we need to evaluate the self-energy functions $\Omega^{(1)},\Omega^{(2)}$ and $\Xi^{(1)}$.
\section{Minimal gauge mediation}
\subsection{Lagrangian of minimal gauge mediation}
The chiral superfields contained in the MGM are messengers $\Phi$ and $\tilde\Phi$,
and a Goldstino multiplet $X$.
The chiral superfield components corresponding to the chiral messenger field $\Phi$
and $\tilde\Phi$, in the fundamental and anti-fundamental representation
of the $SU(3)_c$ gauge symmetry, are denoted as
\begin{equation}
\Phi=(\phi,\psi,{\cal F}),\qquad
\tilde\Phi=(\tilde\phi,\tilde\psi,\tilde{\cal F}).
\end{equation}
The free lagrangian of the messenger fields and the SUSY Yang-Mills lagrangian
are given as follows:
\begin{equation}
\begin{aligned}\label{eq:longlag}
{\cal L}_{\rm free}&= -\frac{1}{4}v^a_{\mu\nu}v^{a\mu\nu}
+i\lambda^{\dagger a}\bar\sigma^\mu D_\mu\lambda^a+\frac{1}{2}{\cal D}^a{\cal D}^a\\
&\quad+(D_\mu \phi)^\dagger (D^\mu \phi)+i\psi^\dagger\bar\sigma^\mu (D_\mu\psi)+{\cal F}^\dagger {\cal F}
+(D_\mu\tilde\phi) (D^\mu \tilde\phi)^\dagger+i\tilde\psi\sigma^\mu (D_\mu\tilde\psi^\dagger)
+\tilde {\cal F} \tilde {\cal F}^\dagger\\
&\quad-\sqrt{2}g(\phi^\dagger {\boldsymbol T}^a\psi - \tilde \psi {\boldsymbol T}^a \tilde\phi^\dagger)\lambda^a
-\sqrt{2}g\lambda^{a \dagger}(\psi^\dagger {\boldsymbol T}^a \phi - \tilde\phi {\boldsymbol T}^a\tilde\psi^\dagger)\\
&\quad+g{\cal D}^a(\phi^\dagger {\boldsymbol T}^a\phi-\tilde \phi {\boldsymbol T}^a\tilde \phi^\dagger),
\end{aligned}
\end{equation}
where (i) $g$ is the gauge coupling, (ii) $f^{abc}$ are the antisymmetric structure constants of the gauge symmetry
which satisfy
\begin{equation}
[{\boldsymbol T}^a,{\boldsymbol T}^b]=if^{abc} {\boldsymbol T}^c,
\end{equation}
for the generators ${\boldsymbol T}^a$ for the fundamental representation,
(iii) $\lambda^a$ is the gluino field,
(iv) $v^a_{\mu\nu}$ the gluon field strength,
\begin{equation}
v^a_{\mu\nu}=\partial_\mu v^a_\nu-\partial_\nu v^a_\mu-gf^{abc}v^b_\mu v^c_\nu,
\end{equation}
(v) ${\cal D}^a$ is the real auxiliary boson field,
and (vi) the covariant derivatives are defined as
\begin{equation}
\begin{aligned}
(D_\mu \phi)_i&= \partial_\mu \phi_i+igv^a_\mu ({\boldsymbol T}^a)_i^{\,\, j} \phi_j,\\
(D_\mu \tilde \phi)^i & = \partial_\mu \tilde\phi^i-ig v^a_\mu \tilde\phi^j ({\boldsymbol T}^a)_j^{\,\, i}\\
(D_\mu \tilde \phi)^\dagger_i &= \partial_\mu \tilde \phi^\dagger_{\,i}
+ igv^a_\mu ({\boldsymbol T}^a)_i^{\,\, j}\tilde \phi^\dagger_{\,j},\\
(D_\mu \psi)_i &= \partial_\mu\psi_i+igv^a_\mu({\boldsymbol T}^a)_i^{\,\, j} \psi_j,\\
(D_\mu \tilde\psi^\dagger)_i &= \partial_\mu\tilde\psi^\dagger_{\,i}
+igv^a_\mu ({\boldsymbol T}^a)_i^{\,\, j}\tilde\psi^\dagger_{\,j},\\
D_\mu \lambda^a&= \partial_\mu\lambda^a-gf^{abc}v^b_\mu\lambda^c.
\end{aligned}
\end{equation}

The Goldstino $X$ couples to the messengers via a superpotential
\begin{equation}
W= X\tilde\Phi\Phi,
\end{equation}
and has an expectation value:
\begin{equation}
\langle X\rangle = M+\theta^2 F_X.
\end{equation}
Its expectation value $F_X$ sets the scale of SUSY breaking as $\sqrt{F_X}$.
The other expectation value $M$ gives each messenger fermion a mass $M$,
and the scalars mass squared masses equal to $M^2\pm F_X$.
The corresponding mass eigenstates of the scalars are given in terms of
the gauge eigenstates of the scalars as follows:
\begin{equation}
\left(\begin{array}{c}\phi_- \\ \phi_+ \end{array}\right)
=\frac{1}{\sqrt{2}}\left(\begin{array}{cc} 1 & 1 \\ 1 & -1 \end{array}\right)
\left(\begin{array}{c}\phi \\ \tilde\phi^\dagger\end{array}\right).
\end{equation}
For the Feynman rules we select the mass eigenstates for the messenger scalars.

\subsection{Feynman rules}
\begin{figure}[t]
\centering
\includegraphics[width=6in]{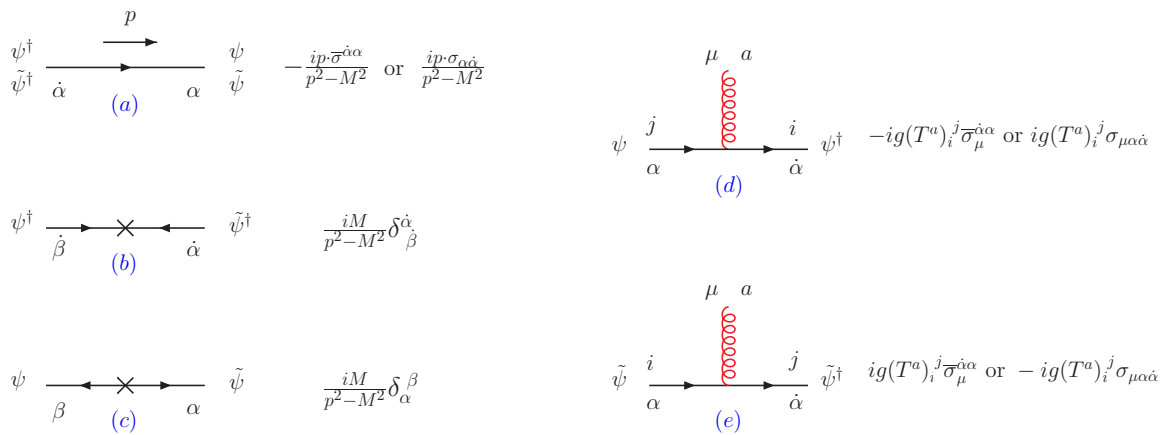}
\caption{The Feynman rules for the messenger fermions}
\label{fig:mfermion}
\end{figure}

\begin{figure}[t]
\centering
\includegraphics[width=5in]{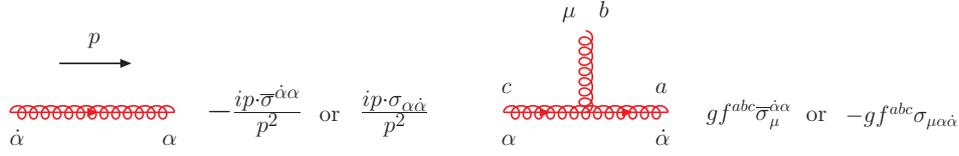}
\caption{The Feynman rules for the gluino}
\label{fig:gluino}
\end{figure}

\begin{figure}[t]
\centering
\includegraphics[width=6 in]{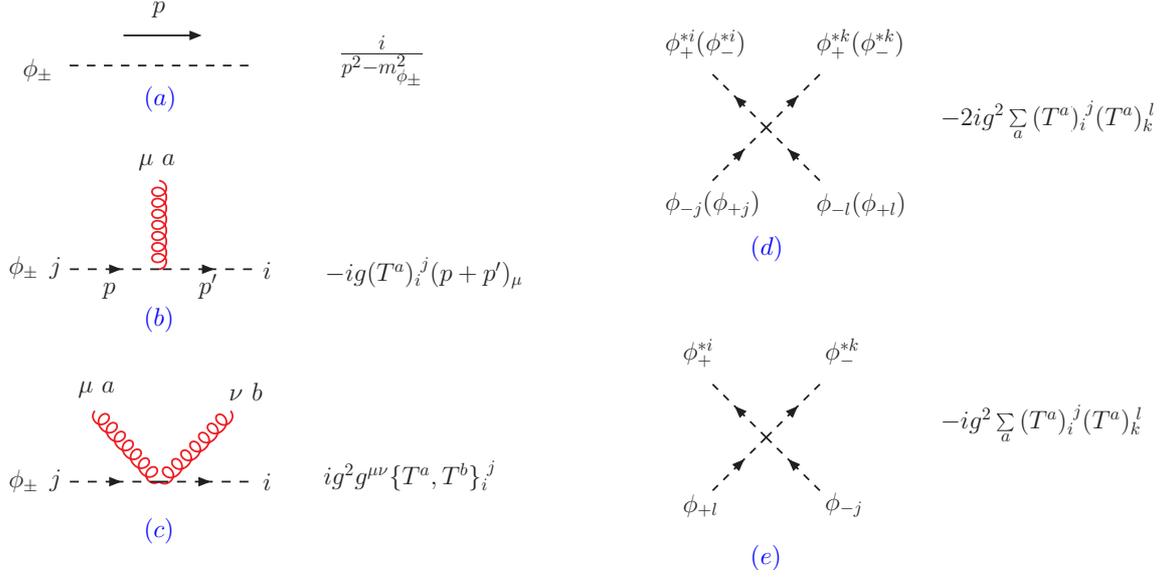}
\caption{The Feynman rules for the messenger scalars}
\label{fig:mscalar}
\end{figure}

\begin{figure}[t]
\centering
\includegraphics[width=5 in]{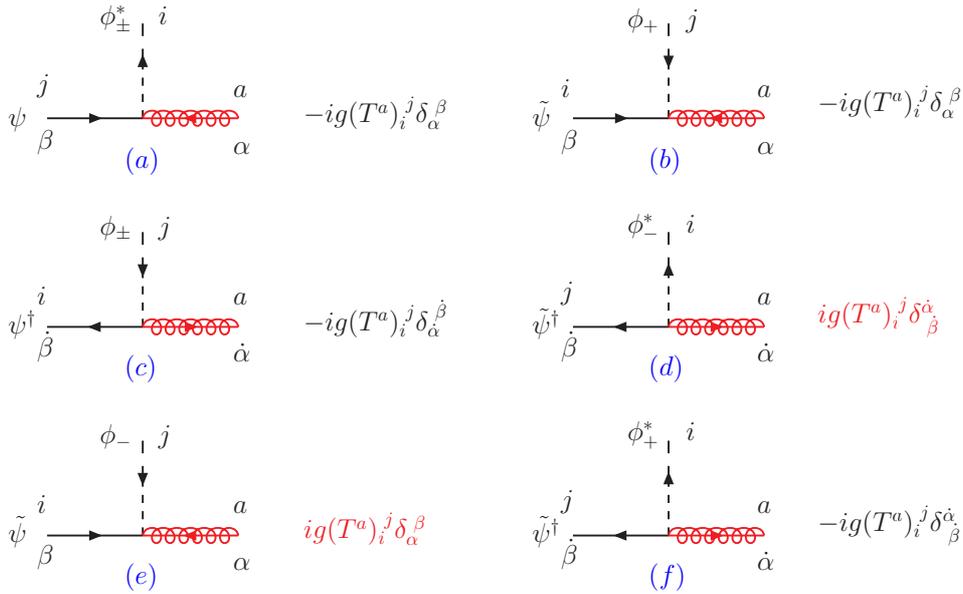}
\caption{The Feynman rules for the scalar-fermion-gluino vertices.}
\label{fig:3vertex}
\end{figure}

In order to systematically perform perturbative calculations
we first establish a set of Feynman rules for the MGM using the two-component spinor formalism.
Following the conventions for the Feynman rules in ref.~\cite{Dreiner:2008tw}
we acquire them as depicted in figure~\ref{fig:mfermion},~\ref{fig:gluino},~\ref{fig:mscalar}, and~\ref{fig:3vertex}.
Here we use the usual Feynman gauge for the gluon field.
We omit the Feynman rules for the gluon for the sake of saving space.

The messenger fermions have four different propagators:
the two are chirality-preserving as shown in figure~\ref{fig:mfermion} (a)
while the other two chirality-violating as shown in figure~\ref{fig:mfermion} (b) and (c).
For the Feynman rules for chirality-preserving propagators we have two options for them:
either $\overline{\sigma}$ or $\sigma$.
For the Feynman rules for fermion-fermion-gluon vertices we can also select either $\overline{\sigma}$
or $\sigma$ as shown in figure~\ref{fig:mfermion}~(d) and (e).
But the choice on $\sigma$ matrices for propagators and the vertices in a Feynman diagram should be simultaneously fixed.
For instance, if one chooses a $\sigma$ for a propagator
then its neighboring vertices must pick out a $\overline{\sigma}$.

As for a gluino we have only a chirality-preserving propagator as shown in figure~\ref{fig:gluino}
because it is massless.
The gluino propagator as well as the gluino-gluino-gluon vertex can contain either $\overline{\sigma}$ or
$\sigma$ like the messenger fermions.

The messenger scalars have two different four-vertices which are obtained by integrating out ${\cal D}^a$-term in Eq.~(\ref{eq:longlag}): the same mass eigenstates have {\it either} the same directions of their arrows as shown in figure~\ref{fig:mscalar}~(d)
{\it or} the opposite directions of their arrows as shown in figure~\ref{fig:mscalar}~(e).

There are eight different scalar-fermion-gluino three-vertices as shown in figure~\ref{fig:3vertex}.
One should pay attention to every direction of arrows of fields in the vertex diagrams.
There are some important properties to recall:
\begin{itemize}
\item A direction of arrow of a messenger fermion ($\psi$ or $\tilde\psi$) is the same as that of gluino:
either into a vertex or out of a vertex.
\item A direction of arrow of a messenger scalar is {\it either}
the same with that of a messenger fermion $\tilde\psi$
{\it or} the opposite with that of the messenger fermion $\psi$.
\item Among the eight vertices only the two in figure~\ref{fig:3vertex}~(d) and (f) have a positive sign
in the Feynman rules while the rest a negative sign.
\end{itemize}

Although we do not include the MSSM quarks and squarks in the lagrangian~(\ref{eq:longlag})
we need to take account of their effects on the renormalization procedure for later numerical analysis.
Here we omit their Feynman rules which are referred to in ref.~\cite{Dreiner:2008tw}.

\section{Renormalization}\label{sect:renor}
\begin{figure}[t]
\centering
\includegraphics[width=6in]{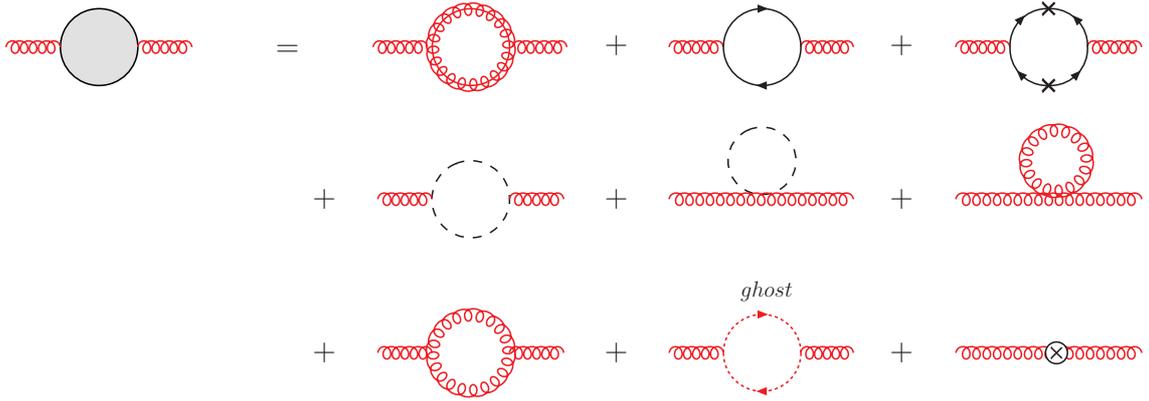}
\caption{The one-loop and counterterm corrections to the gluon propagator.}
\label{fig:renor1}
\end{figure}

\begin{figure}[t]
\centering
\includegraphics[width=6in]{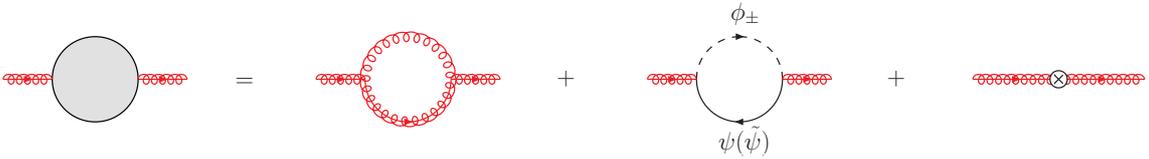}
\caption{The one-loop and counterterm corrections to the chirality-preserving gluino propagator.}
\label{fig:renor2}
\end{figure}

\begin{figure}[t]
\centering
\includegraphics[width=6.0in]{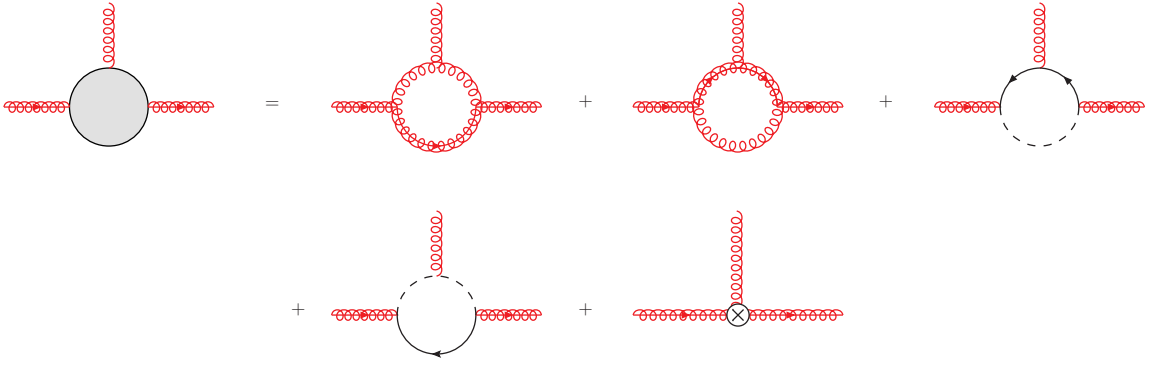}
\caption{The one-loop and counterterm corrections to the gluino-gluino-gluon vertex.}
\label{fig:renor8}
\end{figure}

We shall perform loop calculations for the self-energy and vertex functions
using the Feynman rules as described in the previous section.
We consider the one-loop and counterterm corrections to the propagators
and vertices relevant for the gluino pole mass.
As mentioned earlier we take into account the MSSM quarks and squarks
whose propagators appear in the loops of figure~\ref{fig:renor1},~\ref{fig:renor2}, and~\ref{fig:renor8}.

\begin{figure}[t]
\centering
\includegraphics[width=2.7in]{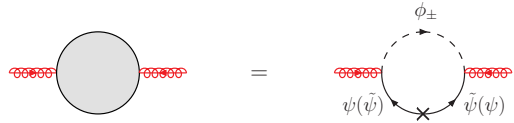}
\caption{The one-loop corrections to the chirality-violating gluino propagator.
They contributes to the pole mass of the gluino at one-loop order.}
\label{fig:renor3}
\end{figure}

\begin{figure}[t]
\centering
\includegraphics[width=5.5in]{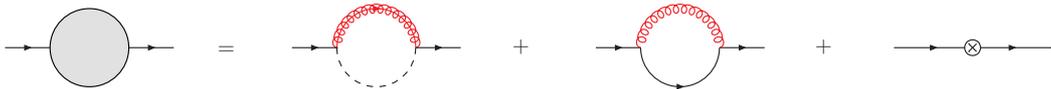}
\caption{The one-loop and counterterm corrections to the chirality-preserving  messenger fermion propagator.}
\label{fig:renor4}
\end{figure}

\begin{figure}[t]
\centering
\includegraphics[width=4.2in]{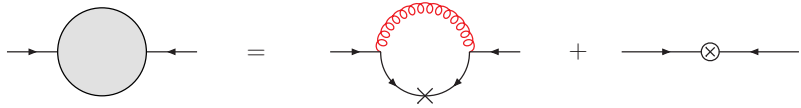}
\caption{The one-loop and counterterm corrections to the  chirality-violating messenger fermion propagator.}
\label{fig:renor5}
\end{figure}

\begin{figure}[t]
\centering
\includegraphics[width=6.0in]{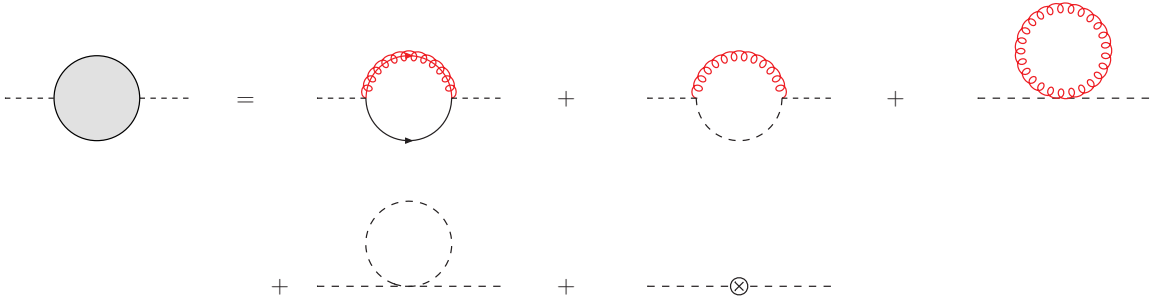}
\caption{The one-loop and counterterm corrections to the messenger scalar propagator.}
\label{fig:renor6}
\end{figure}

All the corrections for the propagators are depicted
in figure~\ref{fig:renor1},~\ref{fig:renor2},~\ref{fig:renor3},~\ref{fig:renor4},~\ref{fig:renor5}, and~\ref{fig:renor6}:
\begin{itemize}
\item Figure~\ref{fig:renor1} shows the one-loop and counterterm corrections to the gluino propagator.
The solid lines represent either the messenger fermions or quarks while the dashed lines
either the messenger scalars or squarks.
The second to the last denotes the ghost loop corrections while the last stands for the counterterm.
\item Figure~\ref{fig:renor2}  shows the one-loop and counterterm corrections to the chirality-preserving gluino propagator.
There are four different combinations of the messenger fermions and messenger scalars in the loop and
their contributions are constructive, leading to $UV$ divergence.
We include not only the messenger fermion-scalar loops but also the quark-squark loops.
\item Figure~\ref{fig:renor3}  shows the one-loop corrections to the chirality-violating gluino propagator.
There are four different combinations of the messenger fermions and messenger scalars in the loop but
their contributions are destructive, resulting in no counterterm.
They yield the pole mass of the gluino at one-loop order.
\item Figure~\ref{fig:renor4} shows the one-loop and counterterm corrections to the chirality-preserving messenger fermion propagator.
\item Figure~\ref{fig:renor5} shows the one-loop and counterterm corrections to the chirality-violating messenger fermion propagator.
\item Figure~\ref{fig:renor6} shows the one-loop and counterterm corrections to the messenger scalar propagator.
\end{itemize}

\begin{figure}[t]
\centering
\includegraphics[width=6.0in]{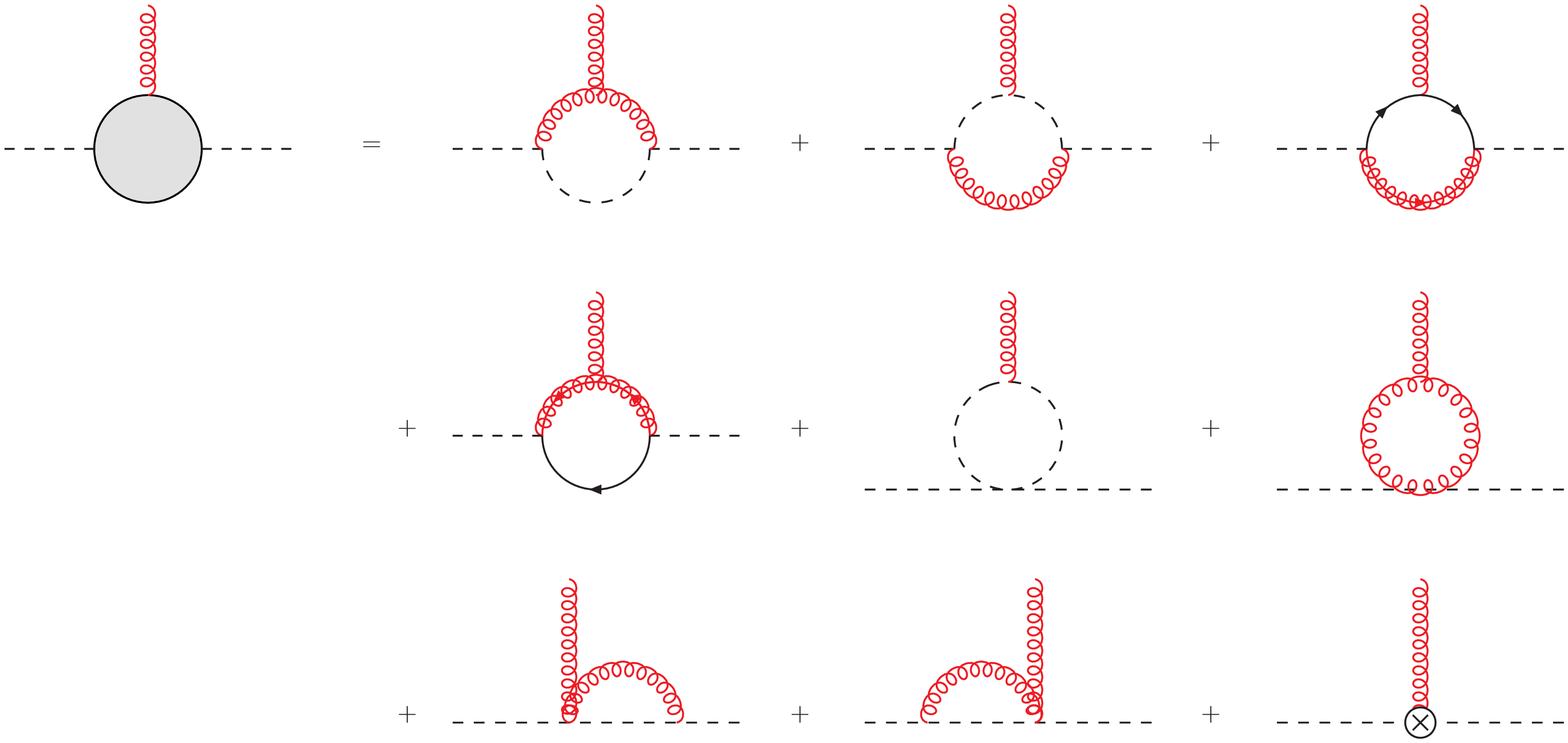}
\caption{The one-loop and counterterm corrections to the messenger scalar-messenger scalar-gluon vertex.}
\label{fig:renor7}
\end{figure}

\begin{figure}[t]
\centering
\includegraphics[width=6.0in]{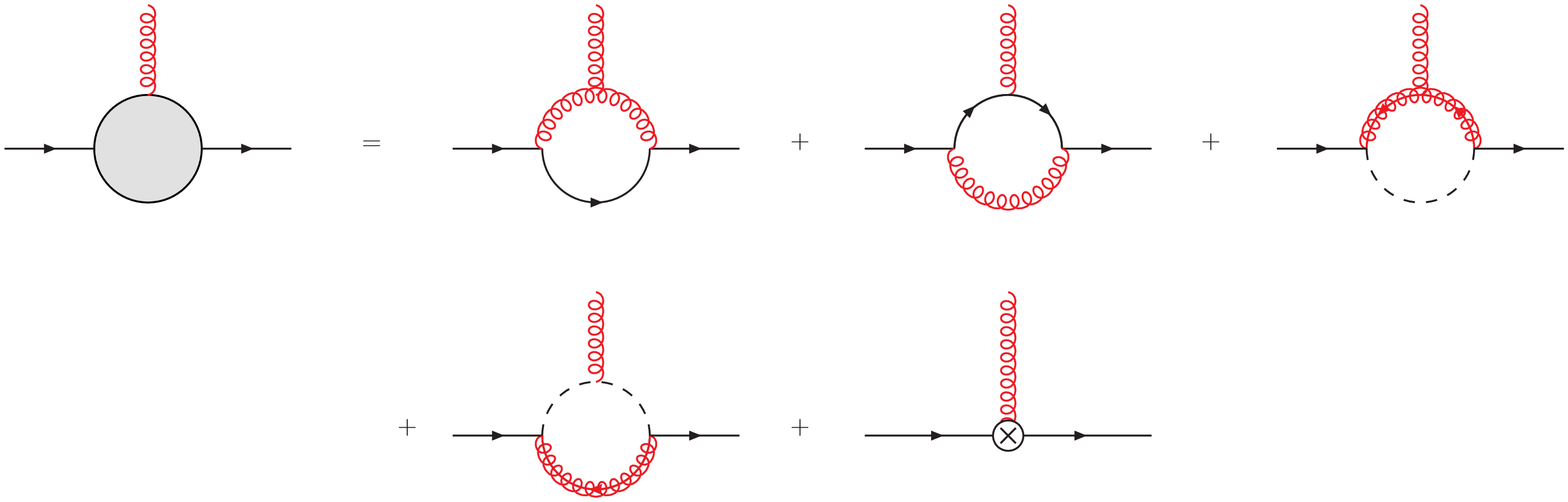}
\caption{The one-loop and counterterm corrections to the messenger fermion-messenger fermion-gluon vertex.}
\label{fig:renor9}
\end{figure}

\begin{figure}[t]
\centering
\includegraphics[width=6.0in]{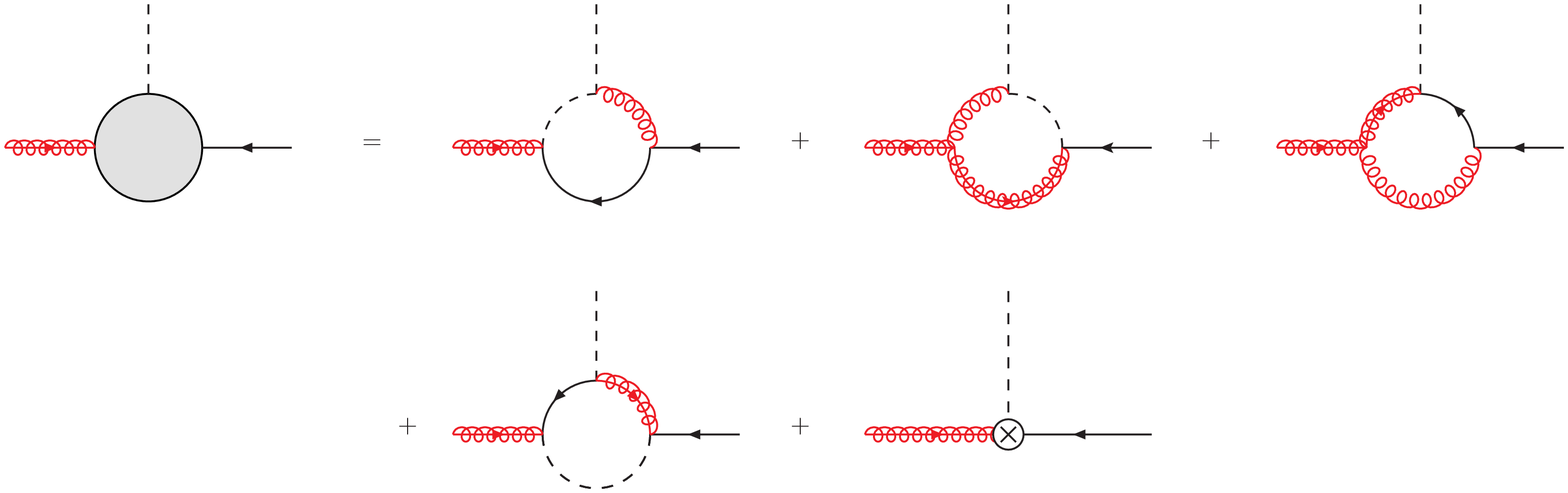}
\caption{The one-loop and counterterm corrections to the messenger scalar-messenger fermion-gluon vertex.}
\label{fig:renor10}
\end{figure}

All the corrections to the three-vertices are depicted
in figure~\ref{fig:renor7},~\ref{fig:renor8},~\ref{fig:renor9}, and~\ref{fig:renor10}:
\begin{itemize}
\item Figure~\ref{fig:renor7} shows the one-loop and counterterm corrections to the messenger scalar-messenger scalar-gluon vertex.
Reversing the arrow direction of either the messenger fermion or gluino is also taken into account.
\item Figure~\ref{fig:renor8} shows the one-loop and counterterm corrections to the gluino-gluino-gluon vertex.
We include not only the messenger fermion-scalar loops but also the quark-squark loops.
\item Figure~\ref{fig:renor9} shows the one-loop and counterterm corrections to the messenger fermion-messenger fermion-gluon vertex.
\item Figure~\ref{fig:renor10} shows the one-loop and counterterm corrections to the messenger fermion-messenger scalar-gluino vertex.
\end{itemize}

The $\overline{\rm{DR}}$ scheme treats the UV divergences in the same way as the $\overline{\rm MS}$ scheme.
The main difference between the two schemes is that degrees of freedom of a chiral fermion field in the $\overline{\rm DR}$ scheme
are set to 2 while those in the $\overline{\rm MS}$ scheme to $d/2$.
Therefore gauge fields in the $\overline{\rm DR}$ scheme are accompanied by the $\varepsilon$-scalar fields to maintain supersymmetry.
At one-loop order, it is equivalent to set the dimensions in which the $\sigma$ matrices reside to 2.

\begin{figure}[t]
\centering
\begin{tabular}{ll}
\epsfig{file= 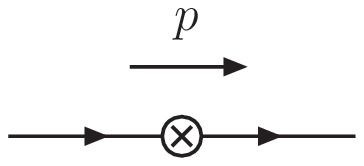, width= 10 ex}  = {\small $ip\cdot \overline{\sigma} (Z_\psi-1)$}, & {\small $Z^{(1)}_\psi=-2C_2(R)$} \\
\epsfig{file= 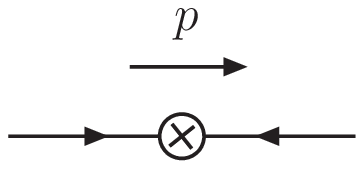, width= 10 ex} = {\small $-iM(Z_\psi Z_M-1)$}, & {\small $Z^{(1)}_M=-2C_2(R)$} \\
\epsfig{file= 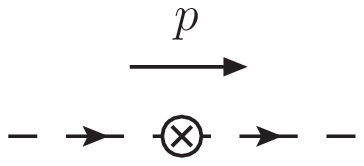, width=10 ex}  = {\small $i[p^2(Z_\phi-1)-M^2(Z_\phi Z^2_M-1)\mp F_X(Z_\phi Z_{F_X}-1)]$}, & {\small $Z^{(1)}_\phi=0,\quad Z^{(1)}_{F_X}=-2C_2(R)$}\\
\epsfig{file= 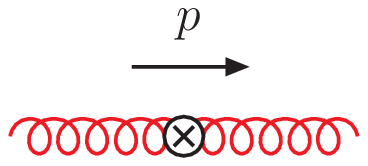, width=10 ex}  = {\small $i(p_\mu p_\nu-p^2g_{\mu\nu})(Z_v-1)$} ,& {\small $Z^{(1)}_v=C_2(G)-2(N_f+ N_{\rm mess})C(R)$}\\
\epsfig{file= 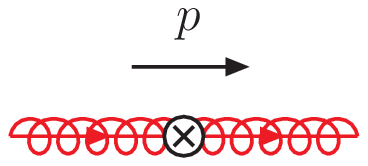, width=10 ex}  = {\small $ip\cdot\overline{\sigma}(Z_\lambda-1)$}, & {\small  $Z^{(1)}_\lambda=-C_2(G)
    -2(N_f+N_{\rm mess})C(R)$} \\
\end{tabular}
\caption{The counterterm corrections to the propagators.}
\label{fig:count1}
\end{figure}

\begin{figure}[t]
\centering
\begin{tabular}{ll}
\epsfig{file= 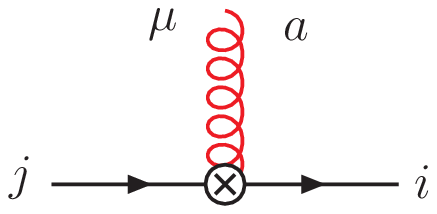, width= 15 ex} = {\small $-ig(T^a)^{\,\,j}_i \overline{\sigma}_\mu (Z_{\psi\psi v}-1)$},
& {\small $Z^{(1)}_{\psi\psi v}=-C_2(G)-2C_2(R)$} \\
\epsfig{file= 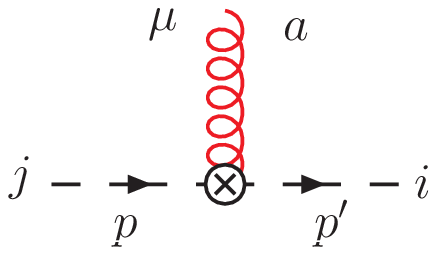, width= 15 ex} = {\small $-ig(T^a)^{\,\,j}_i(p_\mu+p'_\mu)(Z_{\phi\phi v} -1)$},
& {\small $Z^{(1)}_{\phi\phi v}=-C_2(G)$} \\
\epsfig{file= 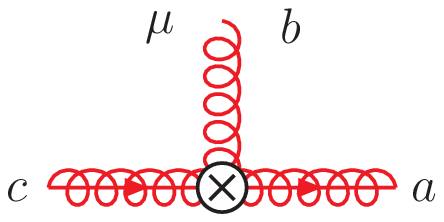, width=15 ex} = {\small $gf^{abc}\overline{\sigma}_\mu (Z_{\lambda\lambda v}-1)$},
& {\small  $Z^{(1)}_{\lambda\lambda v}=-2 C_2(G)-2(N_f+ N_{\rm mess}) C(R)$} \\
\epsfig{file= 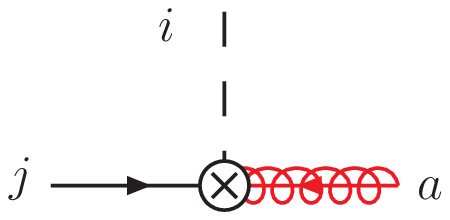, width=15 ex} = {\small $-ig (T^a)^{\,\,j}_i(Z_{\psi\phi\lambda}-1)$},
& {\small $Z^{(1)}_{\psi\phi\lambda}=-2 C_2(G)-C_2(R)$} \\
\end{tabular}
\caption{The counterterm corrections  to the vertices.}
\label{fig:count2}
\end{figure}

The renormalization $Z$ factors are defined as
\begin{equation}
Z=1+\bigg(\frac{\alpha_s}{4\pi}\bigg)\frac{Z^{(1)}}{\varepsilon} +
{\cal O}(\alpha_s^2),
\end{equation}
and cancel off the inverse powers of $\varepsilon$ of the one-loop integrals in the figures.
Both the one-loop corrections to the propagators and vertices yield the leading term of
power series in $\alpha_s$ for $Z^{(1)}$'s.
We list them in figure~\ref{fig:count1} and~\ref{fig:count2} where
$N_f$ is the number of MSSM quark flavor and
$N_{\rm mess}$ is the number of the messenger pairs.
The Casimir operators $C(R)$, $C_2(R)$ and $C_2(G)$ are defined as
\begin{align}
(\boldsymbol{T}^a\boldsymbol{T}^a)^{\,j}_i& =C_2(R)\delta^{\,j}_i\\
\mbox{Tr}[\boldsymbol{T}^a\boldsymbol{T}^b]&=C(R)\delta^{ab}\\
f^{acd}f^{bcd}&=C_2(G)\delta^{ab},
\end{align}
whose values are $C(R)=1/2$, $C_2(R)=4/3$ and $C_2(G)=3$ for the $SU(3)_C$ gauge symmetry, respectively.

As a consistency check we evaluate {\it Slavnov-Taylor} identities,
\begin{equation}
Z_g= \frac{Z_{\psi\psi v}}{Z_\psi \sqrt{Z_v}}=\frac{Z_{\phi\phi v}}{Z_\phi \sqrt{Z_v}}=\frac{Z_{\lambda\lambda v}}{Z_\lambda \sqrt{Z_v}}
=\frac{Z_{\psi\phi v}}{\sqrt{Z_\psi}\sqrt{Z_\phi} \sqrt{Z_v}}=\frac{Z_{vvv}}{\sqrt{Z^3_v}},
\end{equation}
where the various $Z$ factors are calculated at one-loop order.
We obtain the leading coefficient of the renormalization $Z$ factor for the gauge coupling,
\begin{equation}
Z^{(1)}_g=-\frac{3}{2}C_2(G)+(N_f+N_{\rm mess}) C(R).
\end{equation}

\section{Self-energy functions\label{sec:selfe}}

\subsection{Self-energy functions at one-loop order}
We first calculate the helicity-conserving self-energy function $\Omega^{(1)}$ which
arises from the one-loop diagrams as shown in figure~\ref{fig:renor3}.
The sum of the four different configurations in the internal loops is free from UV and IR divergence,
yielding the gluino pole mass at one-loop order as,
\begin{eqnarray}
\sqrt{s}&=&m_{\tilde g}= \Omega^{(1)}(s)\\\nonumber
&=&2 N_{\rm mess} M_{\rm mess} \bigg(\frac{\alpha_s}{4\pi}\bigg) C(R)
\Big(-(\eta_1-1)\ln(\eta_2\,x+1)-(\eta_2-1)\ln(\eta_1\,x+1)
- (r \leftrightarrow \r2 ) \Big),
\end{eqnarray}
where
\begin{align}\label{eq:omega_1}
x &=-\frac{p^2}{M^2_{\rm mess}}=-\frac{s}{M^2_{\rm mess}},\\
r &= \frac{m^2_{\phi_+}}{M_{\rm mess}^2} = 1 + \frac{\Lambda}{M_{\rm mess}},\\
\r2 &= \frac{m^2_{\phi_-}}{M_{\rm mess}^2} = 1 - \frac{\Lambda}{M_{\rm mess}},\\
\eta_{1,2} &= \frac{r+x-1}{2x}\pm\frac{1}{2x}\sqrt{r^2+2r(x-1)+(x+1)^2},
\end{align}
and $p^\mu$ is the external four momentum.
$M_{\rm mess}$ is messenger fermion mass (which is denoted by $M$ in section 3) and
$\Lambda$ is SUSY breaking scale of the visible sector which is equivalent to $F_X/M_{\rm mess}$.
It is noted that the self-energy function $\Omega^{(1)}$ has no explicit dependence on the renormalization scale, $\mu$.
Since the relevant range of $s$ is much lower than the messenger scale, {\it i.e.} $x\leq10^{-4}$,
it is a good approximation to take the limit $x\to0$.
Then $\Omega^{(1)}$ is given by~\cite{Martin:1996zb,Dimopoulos:1996gy},
\begin{equation}
\Omega^{(1)}(s)|_{x \to \,0} =
\bigg(\frac{\alpha_s}{4\pi}\bigg) \Lambda
 N_{\rm mess} \,2C(R) \Bigg[\,\frac{r \ln(r)}{(r-1)^2}
 + (r \leftrightarrow \r2 ) \Bigg].
\end{equation}

The helicity-violating self-energy function $\Xi^{(1)}$, stemming from the diagrams in figure~\ref{fig:renor2}, is given as
\begin{eqnarray}
\Xi^{(1)}(s) &=& \bigg(\frac{\alpha_s}{4\pi}\bigg) C(R) \bigg[
 -N_{\rm mess} \bigg(
 \Big((\eta_1-1)\ln(\eta_2\,x+1)+(\eta_2-1)\ln(\eta_1\,x+1)\Big) \bigg(\frac{r+x-1}{x}\bigg)
 \nonumber \\ && ~~
  +2+\frac{r-1}{x} + \frac{r\ln(r)}{x}
  +\ln\bigg(\frac{\mu^2}{M^2_{\rm mess}}\bigg)
  + (r \leftrightarrow \r2 )
\bigg)
 +2 N_f\bigg(\ln\bigg(\frac{-s}{\mu^2}\bigg)-2\bigg) \bigg]
 \nonumber \\ &&
 +\bigg(\frac{\alpha_s}{4\pi}\bigg)
  C_2(G)\bigg[\ln\bigg(\frac{-s}{\mu^2}\bigg)-2\bigg].
\end{eqnarray}
As mentioned earlier, it consists of the three parts: (i) the messenger loops,
(ii) the quark-squark loops, and (iii) the gluino-gluon loops.
In the limit $x\to 0$, it is reduced to
\begin{eqnarray}
\Xi^{(1)}(s)|_{x\to 0} &=&
\bigg(\frac{\alpha_s}{4\pi}\bigg) \Bigg[
 C(R)\Bigg(  N_{\rm mess}
\bigg( \frac{r\ln(r)}{r-1} + \frac{\r2\ln(\r2)}{\r2-1}
-2 -2\ln\bigg(\frac{\mu^2}{M^2_{\rm mess}}\bigg) \bigg)
 \nonumber \\ &&
 +2 N_f\bigg(\ln\bigg(\frac{-s}{\mu^2}\bigg)-2\bigg)
 \Bigg)
 +C_2(G)\bigg(\ln\bigg(\frac{-s}{\mu^2}\bigg)-2\bigg)
 \Bigg].
\end{eqnarray}
We will comment on the IR behavior in $\Xi^{(1)}(s)$ in the next subsection.

\subsection{Self-energy function at two-loop order}
\begin{figure}[t]
\centering
\includegraphics[width=5in]{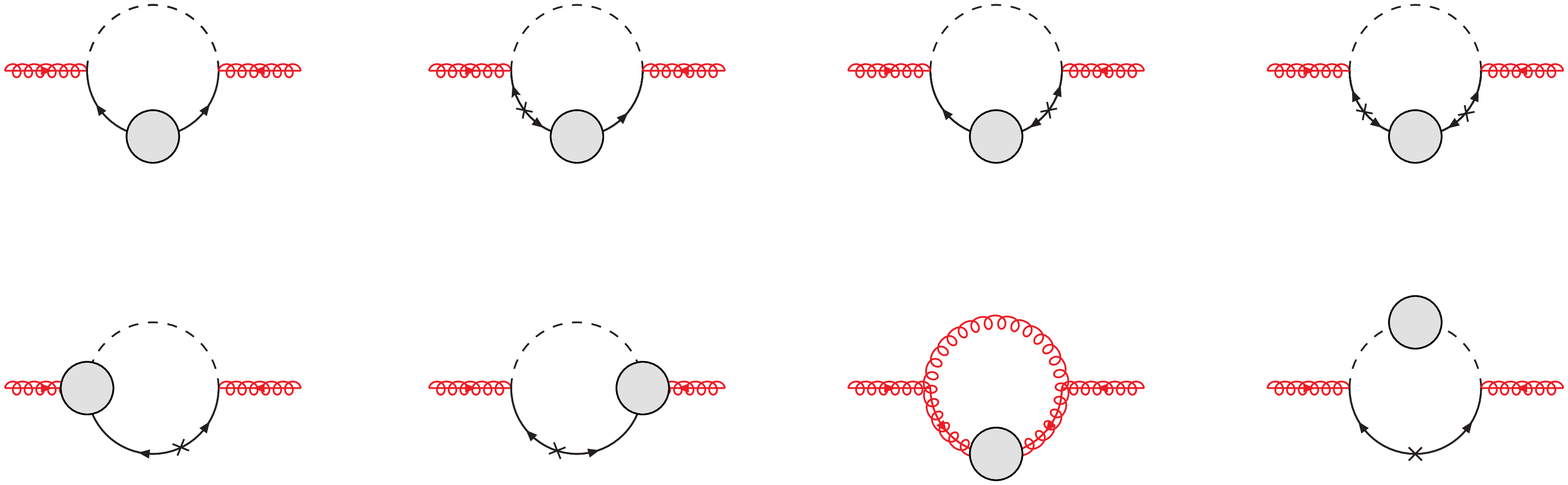}
\caption{The two-loop contributions to the chirality-violating gaugino propagator.
Each shaded circle denotes the one-loop and counterterm
corrections either to the propagators or to the vertices.
The dashed line can be either $\phi_+$ or $\phi_-$ while the solid line either $\psi$ or $\tilde\psi$.
}
\label{fig:twoloop}
\end{figure}
In this section we describe how to calculate the self-energy function $\Omega^{(2)}(s)$ in detail.
All the Feynman diagrams relevant to $\Omega^{(2)}$ are shown in figure~\ref{fig:twoloop}, where
each shaded circle denotes the one-loop and counterterm corrections
either to the propagators or to the vertices in Section~\ref{sect:renor}.
The sum of the Feynman diagrams contains no UV divergence due to the inclusion of the counterterm corrections at one-loop level.

We take a few steps to calculate the Feynman diagrams.
We first decompose a Feynman integral with momentum tensors in numerators
into integral forms of scalar products using the Gram determinant.
Reducing the scalar Feynman integrals to the master integrals we use Laporta's algorithm~\cite{Laporta:2001dd},
which systematically applies several reduction methods,
such as Passarino-Veltman reduction~\cite{Passarino:1978jh},
integration-by-part method~\cite{Tkachov:1981wb,Chetyrkin:1981qh} and
Lorentz invariance method~\cite{Gehrmann:1999as}.
All the reduction procedures have been executed with our in-house Mathematica code.

\begin{figure}[t]
\centering
\includegraphics[width=6.3in]{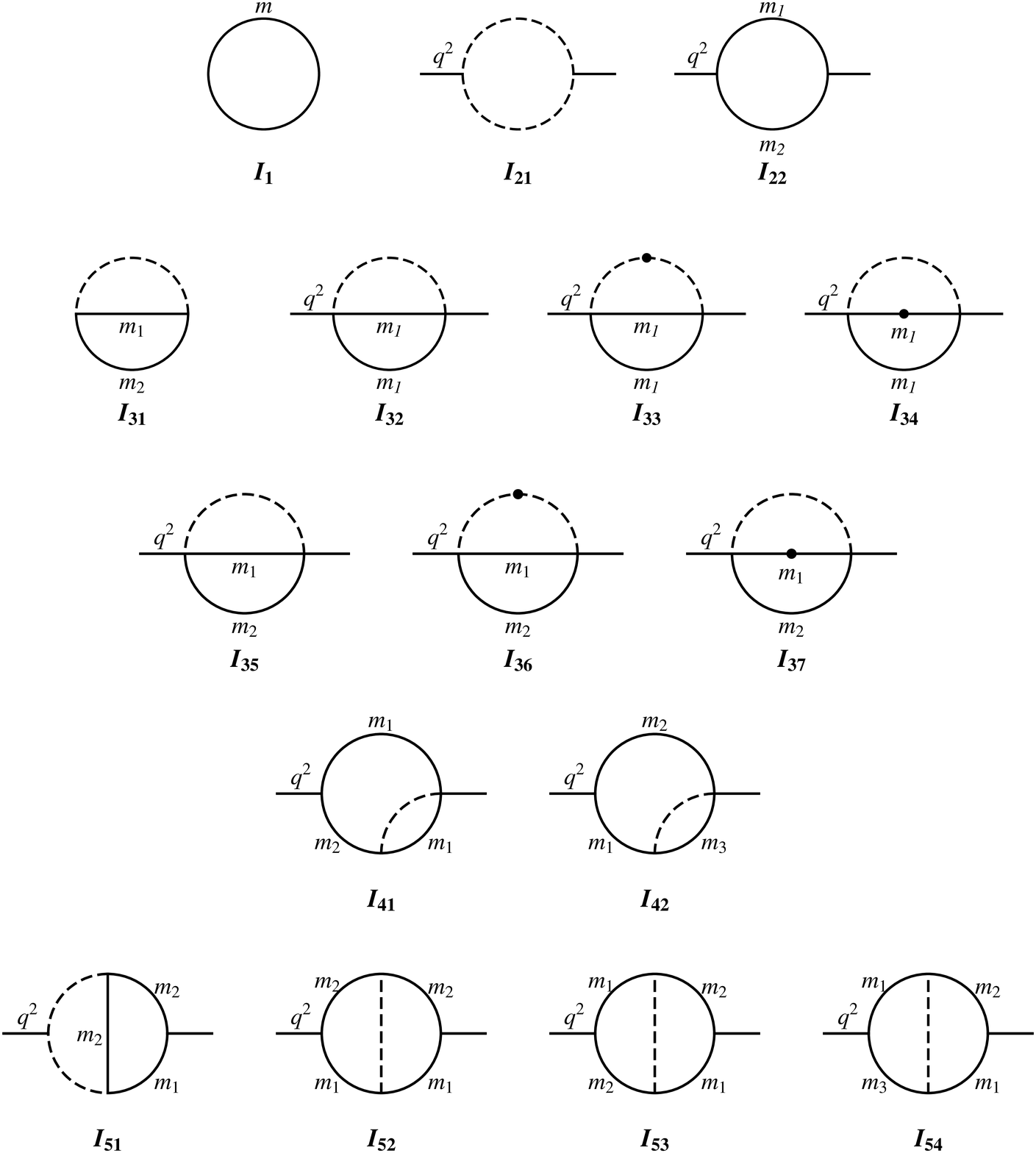}
\caption{The master integrals. A dashed line represents a massless propagator while a solid line a massive propagator
whose mass is explicitly noted. A dot on a solid(dashed) line denotes a double propagator.}
\label{fig:masterIs}
\end{figure}

The {\it sixteen} master integrals are cataloged in figure~\ref{fig:masterIs}.
The dashed lines represent massless propagators while the solid lines massive propagators
whose masses are explicitly noted. A dot on a solid(dashed) line denotes a double propagator.
To evaluate the master integrals we mainly use the Mellin-Barnes representation
which is described in ref.~\cite{Smirnov:2006ry}.
The analytic forms of the master integrals are fully described in appendix~\ref{app:masterint}.

Performing all the procedures we acquire the self-energy function $\Omega^{(2)}(s)$.
We checked that all the UV divergent poles exactly cancel out.
The complete form of $\Omega^{(2)}(s)$ is so lengthy that it is attached in Appendix~\ref{app:fullomega}.
In the limit $x\to0$, the explicit formula of $\Omega^{(2)}$ is approximated by,
{\allowdisplaybreaks
\begin{align}\label{eq:simomega2}
\Omega^{(2)}(s)|_{x\to 0} &=
\bigg(\frac{\alpha_s}{4\pi}\bigg)^2 \Lambda
 N_{\rm mess} \,2C(R) \Bigg[\,
C_2(G)\Bigg( \frac{\ln(r)}{(r-1)^2}\bigg(
8r-4r\ln\Big(\frac{-s}{\mu^2}\Big) \nonumber\\
&+ \Big(2r+\frac{1}{r-1}\Big)\ln(r)
+ \ln(\r2)
\bigg)
+ \frac{4r-2}{(r-1)^2} \,\Li_2(1-r)
+ \frac{2}{r-1}\,\Li_2\bigg(\frac{2r-2}{r}\bigg)
\Bigg)
\nonumber \\
&+C_2(R)\frac{\ln(r)}{(r-1)^2}\Bigg(
6r+2
+2(r+1)\ln\Big(\frac{\mu^2}{M^2_{\rm mess}}\Big)
-\frac{r(r-3)}{r-1}\ln(r)
+\ln(\r2)
\Bigg)
\nonumber \\
& + (r \leftrightarrow \r2 )\Bigg]\,.
\end{align}
}
The first three diagrams (from left to right) on the second row shown in figure~\ref{fig:twoloop}
contribute to the term proportional to $C_2(G)$ in Eq.~(\ref{eq:simomega2})
while all the diagrams except for the one at the intersection of the second row and third column
contribute to the term proportional to $C_2(R)$ in Eq.~(\ref{eq:simomega2}).
The term proportional to $C_2(G)$ contains a term  that represents IR behavior, {\it i.e.} $\ln(-s/\mu^2)$,
since the gluon and gluino in the propagators are massless.
On the other hand, the term proportional to $C_2(R)$ have an large logarithm, $\ln(\mu^2/M^2_{\rm mess})$,
since we take the renormalization scale $\mu$ as the physical gluino mass scale.

As a side note we comment on the large logarithm, $\ln(\mu^2/M^2_{\rm mess})$,
in both $\Xi^{(1)}(s)$ and $\Omega^{(2)}(s)$.
For $N_{\rm{mess}} = 1$ the term with $\ln(\mu^2/M^2_{\rm mess})$ in $\Xi^{(1)}(s)$
gives less than 0.2  even though the $M_{\rm{mess}}$ goes up to $10^5\,\rm{TeV}$
due to the small coupling constant $\alpha_s/(4\pi)=0.008$ at $\mu\sim 1\,\rm{TeV}$.
Therefore we can safely do perturbation for the computation of the pole mass.
However, this term increases linearly with $N_{\rm{mess}}$. For the case $N_{\rm{mess}} \ge 3$ with large $M_{\rm{mess}}$,
it is required to resum this
large logarithmic term in order to make perturbative expansion valid.
As for the $\ln(\mu^2/M^2_{\rm mess})$ term in $\Omega^{(2)}(s)$,
we emphasize that the coefficient of this logarithm turns out to be much smaller than the other terms for a
broad range of the value $r$. Especially in the large messenger mass limit this term vanishes.
Therefore the resummation of the large logarithm is expected to be negligible in $\Omega^{(2)}$.

In order to investigate characteristics of the self-energy functions we consider their behavior 
in a large messenger mass limit ($x\to 0$ and $M_{\rm mess} \gg \sqrt{F_X}$). In this limit,
they become
\begin{eqnarray}
\label{eq:Omega1lim2}
\Omega^{(1)}(s) &=&
\bigg(\frac{\alpha_s}{4\pi}\bigg) \Lambda
 N_{\rm mess} \,2C(R) \,, \\
\label{eq:Omega2lim2}
\Omega^{(2)}(s) &=&
\bigg(\frac{\alpha_s}{4\pi}\bigg)^2 \Lambda
 N_{\rm mess} \,2C(R)
 C_2(G)\bigg(9-4\ln\Big(\frac{-s}{\mu^2}\Big) \bigg) \,, \\
\Xi^{(1)}(s) &=&
\bigg(\frac{\alpha_s}{4\pi}\bigg)
\Bigg[
C(R)\Bigg(
 2 N_f\bigg(\ln\bigg(\frac{-s}{\mu^2}\bigg)-2\bigg)\nonumber\\
&& -2 N_{\rm{mess}} \ln\bigg(\frac{\mu^2}{M^2_{\rm mess}}\bigg)
  \Bigg)
 +C_2(G)\Bigg(
 \ln\bigg(\frac{-s}{\mu^2}\bigg)-2
 \Bigg)
 \Bigg]\,.
\end{eqnarray}
Eq.~(\ref{eq:Omega1lim2}) is the well known result used in various literatures for the gluino mass at one-loop order.
As for $\Omega^{(2)}$ the large logarithm term does not appear
in Eq.~(\ref{eq:Omega2lim2}) as mentioned above. It should be noted that the factor 9
in Eq.~(\ref{eq:Omega2lim2}) is much greater than 2 which are shown in Eq.~(12) of ref.~\cite{Picariello:1998dy}.
It turns out that the difference between the factor 9 and 2 originates from different treatments of IR behavior.
This large factor, as shown in the next section, leads to a significant enhancement to the gluino pole mass
at two-loop order compared with the result in ref.~\cite{Picariello:1998dy}.
We retain an IR behavior in the self-energy functions at two-loop order
while the authors in ref.~\cite{Picariello:1998dy} get rid of an IR divergence.
Since the IR behavior persists in the self-energy functions $\Omega^{(2)}$ and $\Xi^{(1)}$
it is appropriate to keep an IR dependence on the gluino pole mass at two-loop order.
In this regards, our method for the gluino pole mass at two-loop order is more consistent with
perturbation theory rather than that of ref.~\cite{Picariello:1998dy}.

\section{Numerical Analysis}
In this section we perform a numerical analysis of the gluino pole mass using the three self-energy functions
in Section~\ref{sec:selfe}. In order to illustrate the numerical significance of the NLO correction to the gluino pole mass
we compare the NLO pole mass with the LO pole mass.
The strong coupling constant within the standard model is given as $\alpha_s(m_Z)=0.118$,
and the $Z$-boson mass is given as $m_Z=91.187$ GeV, and the top quark mass is given by $m_t=173.1$ GeV.
The top quark mass is required for the $\alpha_s$ running.
The final $\overline{\rm DR}$ values of the gauge coupling $\alpha_s$ and of the fermion masses can be converted into
the $\overline{\rm MS}$ ones at one-loop order~\footnote{The relations between $\overline{\rm DR}$ and $\overline{\rm MS}$
schemes at two- and three-loop order are given in ref.~\cite{Harlander:2006rj} and~\cite{Harlander:2006xq}.} using
\begin{align}
\alpha^{\overline{\rm MS}}_s &=  \alpha^{\overline{\rm DR}}_s\bigg(1-\frac{\alpha_s^{\overline{\rm DR}}}{4\pi}\frac{C_2(G)}{3}\bigg),\\
M^{\overline{\rm MS}} &=  M^{\overline{\rm DR}}\bigg(1+\frac{\alpha_s^{\overline{\rm DR}}}{4\pi}C_2(G)\bigg).
\end{align}

\begin{figure}[t]
\begin{center}
\input{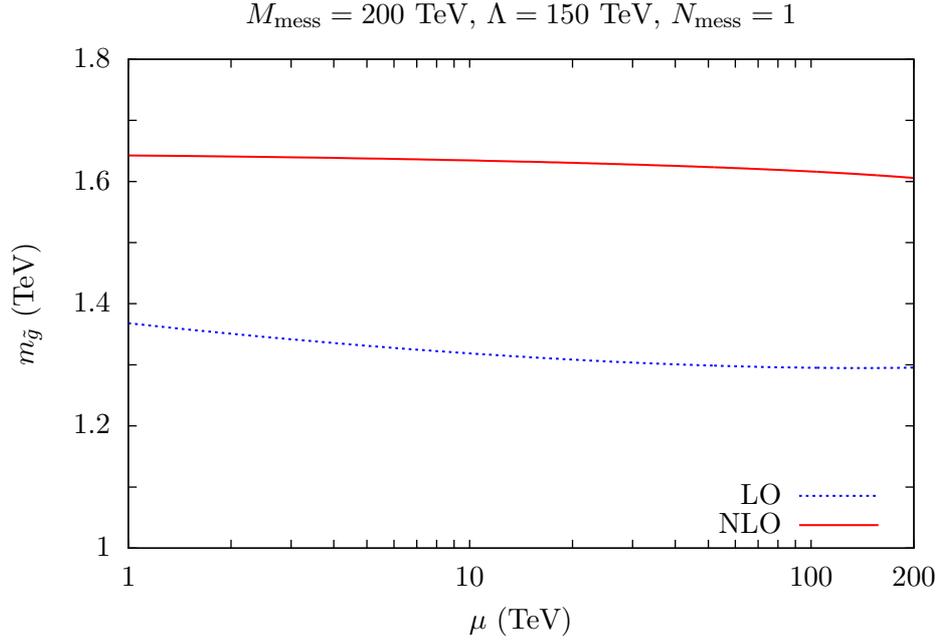}
\caption{The renormalization scale-dependence of the LO (dashed line) and NLO (solid line) pole masses of the gluino
for $\Lambda=150$ TeV, $M_{\rm mess}=200$ TeV and $N_{\rm mess}=1$.}
\label{fig:plot3}
\end{center}
\end{figure}

Figure~\ref{fig:plot3} shows that the renormalization-scale dependence of the LO and NLO pole mass of the gluino,
for $\Lambda=150$ TeV, $M_{\rm mess}=200$ TeV and $N_{\rm mess}=1$.
The LO contribution to the pole mass has no explicit dependence on the renormalization scale $\mu$.
But the introduction of the running gauge coupling brings about the renormalization-scale dependence on the LO pole mass.
Thus the LO pole mass decreases as the renormalization scale increases.
The scale dependence of gluino pole mass is alleviated at the NLO.

\begin{figure}[t]
\begin{center}
\input{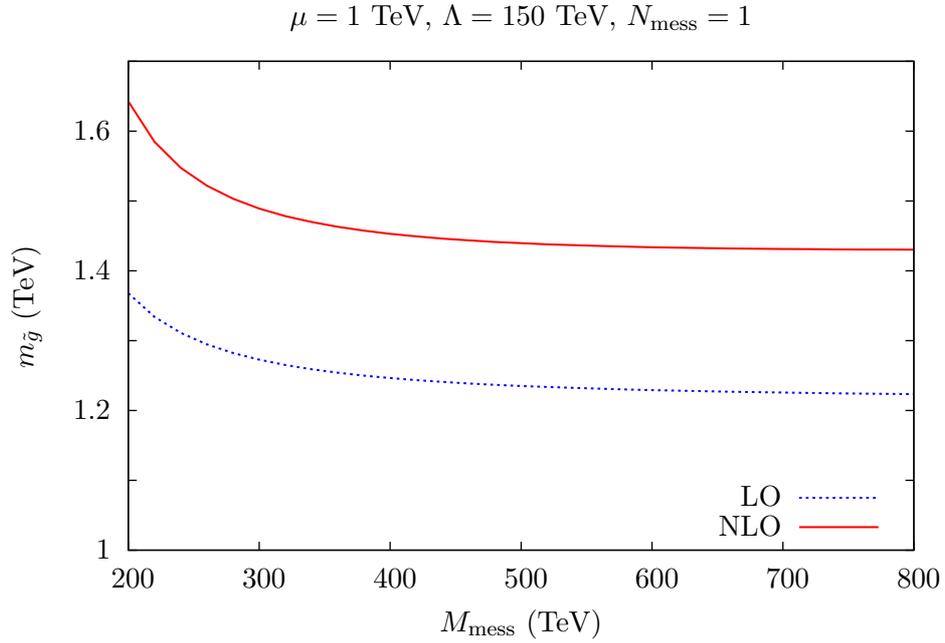}
\caption{The messenger scale-dependence of the LO (dashed line) and NLO (solid line) pole masses of the gluino
for $\mu=1$ TeV, $\Lambda=150$ TeV and $N_{\rm mess}=1$.}
\label{fig:plot1x}
\end{center}
\end{figure}

Figure~\ref{fig:plot1x} compares the LO and NLO pole mass of the gluino as a function of the messenger mass $M_{\rm mess}$,
for $\mu=1$ TeV, $\Lambda=150$ TeV and $N_{\rm mess}=1$.
For a fixed visible supersymmetry breaking scale, both the LO and NLO pole masses are saturated
and the NLO correction barely changes as the messenger mass scale increases.

\begin{figure}[t]
\begin{center}
\input{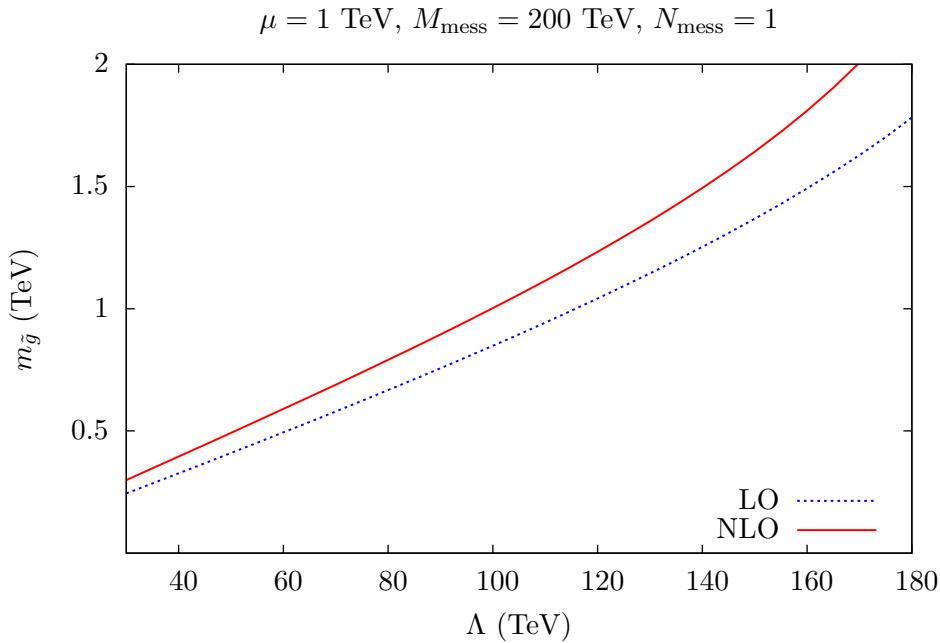}
\caption{The SUSY breaking scale-dependence of the LO (dashed line) and NLO (solid line) pole mass
for $\mu=1$ TeV, $M_{\rm mess}=200$ TeV and $N_{\rm mess}=1$.}
\label{fig:plot2}
\end{center}
\end{figure}

Figure~\ref{fig:plot2} shows both the LO  and NLO pole masses increase
as the visible supersymmetry breaking scale increases for a fixed messenger mass scale.
The NLO correction to pole mass also monotonically increases as $M_{\rm mess}$ increases for fixed values of $\mu$ and $\Lambda$.
All in all, the NLO correction to pole mass is roughly 20\% of the LO  pole masses of the gluino among the three plots.

We consider benchmark planes, lines and points of the MGM
in order to interpret the experimental results in terms of possible manifestations of SUSY.
The authors in ref.~\cite{AbdusSalam:2011fc} proposed the benchmark planes, lines and points of the mGMSB model,
producing the spectra at specific benchmark points that illustrate different possible experimental signatures.
We adopt their benchmark points and calculate the ratio of the NLO pole mass correction to the LO pole mass.
Table~\ref{ta:benchmark1} shows the benchmark points along the benchmark line, \emph{mGMSB1} which is defined
by $N_{\rm mess}=3, \tan\beta=15, \mu>0,\Lambda =M_{\rm mess}/2$ with $\Delta M_{\rm mess}=10$ TeV.
Table~\ref{ta:benchmark2} lists the benchmark point along the benchmark line, \emph{mGMSB2.1} which is defined by
$N_{\rm mess}=1, \tan\beta=15, \mu>0,\Lambda =0.9M_{\rm mess}$ with $\Delta M_{\rm mess}=10$ TeV.
We see that the ratio of the NLO pole mass correction to the LO pole mass for the gluino can reach 32\% in the \emph{mGMSB1}.
The large ratio compared with that of the \emph{mGMSB2.1} is mainly due to $N_{\rm mess}^2$ in the $\Omega^{(1)}\Xi^{(1)}$.
\begin{table}
\centering
\begin{tabular}{|c||c|c|c|c|}
\hline
Points & $M_{\rm mess}$/TeV & $m^{LO}_{\tilde g}$ &  $m^{NLO}_{\tilde g}$ & $\Delta m_{\tilde g}/m^{LO}_{\tilde g}$  \\
\hline\hline
mGMSB1.1 & 70 & 953 & 1256 & 0.32 \\\hline
mGMSB1.2 & 80 & 1089 & 1436 & 0.32 \\\hline
mGMSB1.3 & 90 & 1225 & 1616 & 0.32 \\\hline
mGMSB1.4 & 100 & 1361 & 1796 & 0.32 \\\hline
mGMSB1.5 & 110 & 1497 & 1977 & 0.32 \\\hline
mGMSB1.6 & 120 & 1633 & 2157 & 0.32 \\\hline
\end{tabular}
\caption{Line mGMSB1: $N_{\rm mess}=3, \tan\beta=15, \mu>0, \Delta M_{\rm mess}=10$ TeV (masses in GeV).}
\label{ta:benchmark1}
\end{table}
\begin{table}
\centering
\begin{tabular}{|c||c|c|c|c|}
\hline
Points & $M_{\rm mess}$/TeV & $m^{LO}_{\tilde g}$ &  $m^{NLO}_{\tilde g}$ & $\Delta m_{\tilde g}/m^{LO}_{\tilde g}$  \\
\hline\hline
mGMSB2.1.1 & 80 & 719 & 917 & 0.275 \\\hline
mGMSB2.1.2 & 90 & 809 & 1031 & 0.275 \\\hline
mGMSB2.1.3 & 100 & 898 & 1145 & 0.275 \\\hline
mGMSB2.1.4 & 110 & 988 & 1259 & 0.274 \\\hline
mGMSB2.1.5 & 120 & 1078 & 1373 & 0.274 \\\hline
mGMSB2.1.6 & 130 & 1168 & 1488 & 0.274 \\\hline
\end{tabular}
\caption{Line mGMSB2.1: $N_{\rm mess}=1, \tan\beta=15, \mu>0, \Lambda=0.9 M_{\rm mess}$ with $\Delta M_{\rm mess}=10$ TeV (masses in GeV). }
\label{ta:benchmark2}
\end{table}

All the numerical analyses indicate that the NLO correction is large enough to reach 20\% of the LO pole mass or even more.
There are three ingredients for the large correction: (i) the relative strength of the $SU(3)_C$ gauge coupling
is large, (ii) the color representation of the gluino is octet,
(iii) and the number of the Feynman diagrams associated with the messenger fields are large.
The first two reasons are same with that of gravity mediation mechanism ({\it i.e.} CMSSM)
while the last is characteristic of gauge mediation mechanism.

\section{Conclusion and Outlook}
We have presented the self-energy functions for
the gluino of the minimal gauge mediation at two-loop order and studied the radiative corrections on the gluino pole mass.
The one-loop pole mass is the leading order while the two-loop correction is the next-to-leading order.
The next-to-leading order correction shifts the leading order pole mass by roughly 20\% or even more.
This shift is much larger than the expected accuracy of the mass determination at the LHC,
and should be reckoned with for precision studies on the SUSY breaking parameters.

Not only the gluino mass but also the squark masses are crucial to study the phenomenology at the LHC.
The squark masses also involve the SUSY QCD so that its contribution from higher order radiative corrections are expected to be large.
The numerical significance of the next-to-leading corrections to the squark masses deserves a detailed investigation
which we leave for future study.

The next-to-lightest-superstmmetric particle (NLSP) in gauge mediation is mostly either neutralino or stau depending
on the specific regions in parameter space. Using the full expressions of the self-energy functions one can
evaluate the significance of the radiative corrections for the NLSP mass and refine the spectra at
the benchmark points of the minimal gauge mediation model.

In this paper we have focused only on the minimal gauge mediation.
Other gauge mediation models tend to retain the feature that the higher order radiative corrections
to the gluino pole mass are substantial. Thus when one quantitatively studies
complicated SUSY-breaking models associated with gauge mediation
one must pay attention to the next-to-leading order or higher order radiative corrections to the gluino mass
in addition to the leading order result.

\acknowledgments
J.Y.~Lee and Y.W.~Yoon are supported by Basic Science Research Program through the National Research Foundation of Korea(NRF)
funded by the Ministry of Education, Science and Technology(2011-0003974).
J.Y.~Lee is also supported partially by Mid-career Research Program through the NRF grant funded by the MEST(2011-0027559).
Y.W.~Yoon thanks KIAS Center for Advanced Computation for providing computing resources.
\appendix




\section{Master Integrals\label{app:masterint}}
Reduction of Feynman integrals to master integrals is performed
in the Euclidean space with $d=4-2\varepsilon$ dimensions.
Our convention for the master integral with $n$-propagators (${\cal P}_1,{\cal P}_2,...,{\cal P}_n$), and $h$-loops is
\begin{equation}
(\mu e^{\gamma_E/2})^{h(4-d)} \int \prod_{j=1}^h \Big( \frac{d^d l_j}{i \pi^{d/2}}\Big)\, \frac{1}{{\cal P}_1{\cal P}_2\cdots{\cal P}_n}\,,
\end{equation}
where $l_j$ are loop momenta.
In order to simplify the form of master integrals, several parameters are defined as follows,
 \begin{align}
 ~~~ x &= -\frac{q^2}{m_1^2}-i0, ~~~  r = \frac{m_2^2}{m_1^2}\,,~~~  {\tilde r} = \frac{m_3^2}{m_1^2}\,,  \nonumber \\
 y &= \frac{ \sqrt{x+4} -\sqrt{x} }{\sqrt{x+4} + \sqrt{x}} \,, \nonumber\\
 \kappa_{1,2} &= \frac{1}{2} \pm \frac{\sqrt{x(x+4)}}{2 x} \,, \nonumber \\
 \eta_{1,2} &= \frac{r+x-1}{2 x} \pm \frac{1}{2 x} \sqrt{r^2+2\,r(x-1)+(x+1)^2} \,,\nonumber\\
 \rho_{1,2} &= \frac{\r2-r+x}{2 x} \pm \frac{1}{2 x} \sqrt{ r^2-2r(\r2-x)+(\r2+x)^2} \,,
 \end{align}
 where the subscript 1(2) is associated with a plus(minus) sign.

The master integrals shown in figure~\ref{fig:masterIs} are given in terms of the $\varepsilon$-expansion as follows,
 {\allowdisplaybreaks
 \begin{align}
 I_1 &= -m^2 \Big(\frac{\mu^2}{m^2}\Big)^{\varepsilon}( e^{\gamma_E \varepsilon})\Gamma(-1+\varepsilon\,) \,, \\
 I_{21} &= \Big(\frac{\mu^2}{-q^2}\Big)^{\varepsilon}( e^{\gamma_E \varepsilon})
 \frac{\Gamma(1-\varepsilon)^2 \Gamma(\varepsilon)}{\Gamma(2-2\epsilon)} \,, \\
  I_{22} &= \Big(\frac{\mu^2}{m_1^2}\Big)^{\varepsilon} \Bigg[
 \frac{1}{\varepsilon} + \left(\eta_2-1\right) \ln \left(\eta_1 \,x+1\right)+
 \left(\eta_1-1\right) \ln \left(\eta_2 \,x+1\right)+2 \nonumber \\
 &~~~~~~~~~~ + J_{22}^{(1)}(x,r)\, \varepsilon + {\cal O} (\varepsilon^2) \Bigg]\,, \\
  I_{31} &= m_1^2 \Big(\frac{\mu^2}{m_1^2}\Big)^{2\varepsilon} \Bigg[
  \frac{(1+r)}{2 \varepsilon^2}
 +\left(\frac{3 (1+r)}{2}-r \ln (r)\right)\frac{1}{\varepsilon }
 + (1-r) \text{Li}_2(1-r) \nonumber \\
 & ~~~~~~~~~+\frac{1}{12} \left(42+\pi^2\right) (r+1)+\frac{1}{2} r \ln^2(r)-3 r
   \ln(r) + {\cal O} (\varepsilon)\Bigg]\,, \\
  I_{32} &= m_1^2 \Big(\frac{\mu^2}{m_1^2}\Big)^{2\varepsilon} \Bigg[
   \frac{1}{\varepsilon^2} +\frac{1}{\varepsilon}\bigg(3+\frac{x}{4} \bigg)
 +6+\frac{\pi^2}{6}+\frac{13}{8}x \nonumber \\
& ~~~~~~~~~  +\bigg(\frac{x}{2}-1\bigg)\frac{1+y}{1-y} \ln y
- \bigg(\frac{1+x}{x}\bigg) \ln^2 y +~ {\cal O}(\epsilon) \Bigg]\,,\\
  I_{33} &= \Big(\frac{\mu^2}{m_1^2}\Big)^{2\varepsilon} \Bigg[
  -\frac{1}{2\varepsilon^2} -\frac{1}{\varepsilon} \bigg(\frac{3}{2}
  + \frac{1+y}{1-y} \ln y \bigg) -\frac{9}{2} -\frac{\pi^2}{12}  \nonumber \\
&~~~~~~~~~~ +\frac{1+y}{1-y}\bigg( 6\Li_2(-y) +2\Li_2(y) - 2\ln^2(y) \nonumber \\
&~~~~~~~~~~ + \big(2\ln(1-y) + 6\ln(1+y)-3\big)\ln y + \frac{\pi^2}{6} \bigg)
 +\frac{(1+x)}{x} \ln^2 y + {\cal O}(\varepsilon) \Bigg]\,,\\
   I_{34} &= \Big(\frac{\mu^2}{m_1^2}\Big)^{2\varepsilon} \Bigg[
  \frac{1}{2\varepsilon^2} +\frac{1}{2\varepsilon} -\frac{1}{2} +\frac{\pi^2}{12}
  -\frac{1+y}{1-y} \ln y 
  -\bigg( \frac{1}{x} + \frac{1}{2} \bigg) \ln^2 y + {\cal O}(\varepsilon) \Bigg] \,, \\
  I_{35} &= m_1^2 \Big(\frac{\mu^2}{m_1^2}\Big)^{2\varepsilon} \Bigg[
  \frac{1+r}{2\,\varepsilon^2}
  +\left(\frac{3(r+1)}{2} +\frac{x}{4}  - r\ln(r)\right) \frac{1}{\varepsilon}
  + J_{35}(x,r) +  J_{35}^{(1)}(x,r)\varepsilon + {\cal O}(\varepsilon^2)\Bigg] \,,\\
  I_{36} &= \Big(\frac{\mu^2}{m_1^2}\Big)^{2\varepsilon} \Bigg[
  -\frac{1}{2 \varepsilon ^2} - \bigg(\left(\eta_2-1\right) \ln \left(\eta _1 \,x+1\right)+\left(\eta_1-1\right)
 \ln \left(\eta_2 \,x+1\right)+\frac{3}{2}\,\bigg)\frac{1}{\varepsilon }
 \nonumber \\
  & ~~~~~~~~~~+ J_{36}(x,r)
  + {\cal O}(\varepsilon) \Bigg]\,,\\
 I_{37} &=  \Big(\frac{\mu^2}{m_1^2}\Big)^{2\varepsilon} \Bigg[
 \frac{1}{2\varepsilon^2} + \frac{1}{2\varepsilon}
 + J_{37}(x,r)  +  J_{37}^{(1)}(x,r) \varepsilon + {\cal O}(\varepsilon^2) \Bigg]\,,\\
  I_{4i} &= \Big(\frac{\mu^2}{m_1^2}\Big)^{2\varepsilon} \Bigg[
 \frac{1}{2\,\varepsilon^2} +
 \left(\left(\eta_2-1\right) \ln \left(\eta _1\, x+1\right)+\left(\eta_1-1\right)
 \ln \left(\eta_2 \,x+1\right)+\frac{5}{2} \right)
 \frac{1}{\varepsilon} \nonumber \\
  &~~~~~~~~~~ + J_{4i}(x,r,\tilde r) + {\cal O}(\varepsilon) \Bigg]\,,\\
  I_{5i} &= \frac{1}{m^2_1} J_{5i}(x,r,\tilde r) +  {\cal O}(\varepsilon)\,.
\end{align}
}
The results of the {\it multiple inverse binomial sums} in ref.~\cite{Davydychev:2003mv}
are used for $I_{32}$, $I_{33}$ and $I_{34}$.

The higher transcendental functions $J_{ij}$ and $J_{ij}^{(1)}$ are, respectively,
the zeroth- and first-order terms in the $\varepsilon$-expansion of the master integrals.
The functions with polylogarithms up to second order are given by,
{\allowdisplaybreaks
 \begin{align}
 \label{eq:J22(1)}
 J_{22}^{(1)}(x,r) &=
 \frac{(-r+x+1)}{4 x}\ln ^2(r)
  +\frac{\pi^2}{12}
  +4
  \nonumber \\ &
 +\Bigg[
  \left(2 \left(\eta _2-1\right)
  +\frac{1}{2}\left(\eta _1-\eta _2\right)
  \ln \left(-\left(\eta_1-\eta_2\right){}^2 x^2\right)\right)
  \ln \left(\eta_1 x+1\right)
 \nonumber \\ & ~~~
 + \left(\eta_2-\eta_1\right) \text{Li}_2\left(\frac{1}{\eta_1}\right)
 +\frac{1}{2} \left(\eta_2-\eta _1\right)
  \text{Li}_2\left(\frac{\eta_2 x+1}{\eta_1 x+1}\right)
 + (\eta_1 \leftrightarrow \eta_2 )
 \Bigg]
 \,, \\
\label{eq:J35}
 J_{35}(x,r) &=
 \left(\frac{r}{2 x}+r\right) \ln^2(r)
 + \bigg(3+\frac{\pi^2}{12} \,\bigg) (r+1)
 +\frac{13 x}{8}
 \nonumber \\ &
 +\Bigg[
  \,\frac{1}{2} \left(\eta_2 (-r+x-1)-6 r-x+1\right) \ln \left(\eta_1\,x+1\right)
 \nonumber \\ & ~~~
 -\frac{r (x+1)}{2 x} \ln^2\left(\eta_1\,x+1\right)
 + (1-r) \text{Li}_2\left(\frac{1}{\eta_1}\right)
 + (\eta_1 \leftrightarrow \eta_2 )
 \Bigg]
 \,,\\
\label{eq:J36}
 J_{36}(x,r) &=
 \bigg(\frac{r}{4 x} -\frac{3 (x+1)}{4x}\bigg) \ln^2(r)
 -\frac{\pi ^2}{12}
 -\frac{9}{2}
 \nonumber \\ &
 +\Bigg[
 3\Bigg(
     1
     - \eta_2
     -\frac{1}{2} \left(\eta_1-\eta_2\right) \ln \left(-\left(\eta_1-\eta _2\right){}^2 x^2\right)\Bigg)
     \ln \left(\eta _1 \,x+1\right)
 \nonumber \\ &
 +\bigg(1-\eta _2+\frac{r}{2 x}\bigg) \ln^2\left(\eta_1 \,x+1\right)
 +\bigg(2-4 \eta_2+\frac{r-1}{x}\bigg)\text{Li}_2\left(\frac{1}{\eta_1}\right)
 \nonumber \\ &
 +\left(\frac{3(r+x-1)}{2 x}-3 \eta_2\right)
 \text{Li}_2\left(\frac{\eta_2 \,x+1}{\eta _1\,x+1}\right)
 + (\eta_1 \leftrightarrow \eta_2 )
 \Bigg]
 \,,\\
 \label{eq:J37}
 J_{37}(x,r) &=
 \frac{r \ln^2(r)}{2 x}
 +\frac{\pi^2}{12}
 -\frac{1}{2}
 +\Bigg[
 -\frac{r }{2x}\ln^2\left(\eta _1 x+1\right)
  \nonumber \\ & ~~~
 +\left(1-\eta _2\right) \ln \left(\eta _1 x+1\right)
 +\text{Li}_2\left(\frac{1}{\eta _1}\right)
 + (\eta_1 \leftrightarrow \eta_2 )
 \Bigg]
  \,,\\
 \label{eq:J41}
 J_{41}(x,r) &=
 \frac{(1-r) }{2x}\ln^2(r)
 -4 \ln (r)
 +\frac{(1-r)}{r} \text{Li}_2(1-r)
 +\frac{19}{2}
 \nonumber \\ &
 -\frac{1}{r x} \ln (-x) \ln (x+1)
 +\pi^2\left(\frac{1}{3 r x}+\frac{1}{12}\right)
 \nonumber \\ &
 + \bigg(1 -2 \kappa _1+\frac{2 \ln \left(-\kappa _1\right)}{r x}
 +\frac{\ln (-x)}{r x}\bigg) \ln \left(\kappa _1 x+1\right)
 \nonumber \\ &
 +\frac{(x-1)}{r x} \ln^2\left(\kappa_1 x+1\right)
 +\frac{(x+1)}{r x}
  \left( \text{Li}_2\bigg(\frac{1}{\kappa _1}\bigg)
       + \text{Li}_2\bigg(\frac{1}{\kappa _2}\bigg) \right)
  \nonumber \\ &
 -\frac{1}{r x}
 \left( \text{Li}_2\bigg(\,\frac{x+1}{\kappa _1 x+1}\bigg)
       +\text{Li}_2\bigg(\,\frac{x+1}{\kappa _2 x+1}\bigg)
       + \text{Li}_2(-x) \right)
 \nonumber \\ &
 + \Biggl[
 \eta_1 \Bigg(
 \frac{\pi ^2}{6}
 +\ln \left(\left(\eta_1-\eta_2\right) x\right) \ln\left(\eta_1 x+1\right)
 \nonumber \\ & ~~~
 +\Big(4+\ln \left(\eta_1 x+1\right)
    -\ln \left(\left(\eta_1-\eta_2\right) x\right)\Big)\ln\left(\eta_2 x+1\right)
 \nonumber \\ & ~~~
  -\text{Li}_2\bigg(\frac{1}{\eta_1}\bigg)
  +\text{Li}_2\bigg(\frac{1}{\eta_2}\bigg)
  -\text{Li}_2\bigg(\,\frac{\eta_2\,x+1}{\eta_1\,x+1}\bigg) \Bigg)
 \nonumber \\ & ~~~
 +\frac{(r-1) \eta_1}{r} \Bigg(
 -\ln \left( \eta_1 (1-r)\right) \ln \left(\eta _2\,x+1\right)
 +\text{Li}_2\bigg(\frac{1}{\eta_1}\bigg)
  \nonumber \\ & ~~~
  +\text{Li}_2\bigg(\,\frac{\eta_1-1}{\eta_1-\kappa_1}\bigg)
  -\text{Li}_2\bigg(\,\frac{\eta_1}{\eta_1-\kappa_1}\bigg)
  +\text{Li}_2\bigg(\,\frac{\eta_1-1}{\eta_1-\kappa_2}\bigg)
 \nonumber \\ & ~~~
  -\text{Li}_2\bigg(\,\frac{\eta_1}{\eta_1-\kappa_2}\bigg)
 \Bigg)
  + (\eta_1 \leftrightarrow \eta_2 )
 \Bigg]
  \,,  ~~~~~ ({\rm for}~ x>-1) \\
 \label{eq:J42}
 J_{42}(x,r,\r2) &=
 -\frac{\left(\tilde{r}-2\right) (1-r+x)}{2x}\ln^2(r)
 -\frac{1}{2}\, \tilde{r} \ln^2(\tilde{r})
 -\rho_1 \ln(r)
 +\left(\rho_1-1\right) \ln \left(\tilde{r}\right)
  \nonumber \\ &
 +\left(1-\tilde{r}\right) \text{Li}_2\left(1-\tilde{r}\right)
 +\frac{\pi ^2}{12}
 +\frac{19}{2}
 +\ln \left(\frac{\rho_1 x}{r}+1\right)
   \left(\frac{2 r \tilde{r}}{x}\ln\left(\frac{r}{\r2}\right)
  -\rho_1+\rho_2\right)
 \nonumber \\ &
 +\frac{2 r \tilde{r} }{x}\ln ^2\left(\frac{\rho_1 x}{r}+1\right)
 -\frac{\tilde{r}(r+x)}{x} \left(\text{Li}_2\bigg(\frac{1}{1-\rho_1}\bigg)
    +\text{Li}_2\bigg(\frac{1}{1-\rho_2}\bigg)\right)
  \nonumber \\ &
 +\frac{r \tilde{r}}{x}
 \left(\text{Li}_2\bigg(\,\frac{1-\rho _1}{\rho_1 \left(\rho_2-1\right)}\bigg)
 +\text{Li}_2\bigg(\,\frac{1-\rho_2}{\rho_2\left(\rho_1-1\right)}\bigg)
 -\text{Li}_2\left(-\frac{x}{r}\right)\right)
 \nonumber \\ &
  + \Bigg[
 (\eta_1-1) \Bigg(
 \frac{\pi ^2}{6}
 +\frac{\ln ^2(r)}{2}
 +\ln \left(\left(\eta_1-\eta_2\right) x\right) \ln \left(\eta _1\,x+1\right)
  \nonumber \\ &  ~~~
 -\ln^2\left(\eta _1 x+1\right)
 + \Big(4 -\ln \left(\tilde{r}\right)
    -\ln \left(\left(\eta_1-\eta_2\right) x\right)\Big)
     \ln \left(\eta _2 x+1\right)
 \nonumber \\ & ~~~
 +\text{Li}_2\bigg(\frac{1}{1-\eta_1}\bigg)
 -\text{Li}_2\bigg(\frac{1}{1-\eta_2}\bigg)
 -\text{Li}_2\bigg(\,\frac{\eta_2\,x+1}{\eta_1\,x+1}\bigg) \Bigg)
 \nonumber \\ &  ~~~
 +(\eta_1-1)(\r2-1) \Bigg(
 -\frac{1}{2} \ln^2(r)
  +\ln \left(\eta_2 x+1\right)
  \ln \bigg(\frac{(1-\tilde{r})\left(\eta_1-1\right)}{\tilde{r}} \bigg)
 \nonumber \\ &  ~~~
 +\text{Li}_2\bigg(\frac{1}{1-\eta_1}\bigg)
 +\text{Li}_2\bigg(\,\frac{\eta_1}{\eta_1+\rho_1-1}\bigg)
 +\text{Li}_2\bigg(\,\frac{\eta_1}{\eta_1+\rho_2-1}\bigg)
 \nonumber \\ &  ~~~
 -\text{Li}_2\bigg(\,\frac{\eta _1-1}{\eta _1+\rho_1-1}\bigg)
 -\text{Li}_2\bigg(\,\frac{\eta _1-1}{\eta _1+\rho_2-1}\bigg)
  \Bigg)
  + (\eta_1 \leftrightarrow \eta_2 )
 \Bigg]
  \,. \\ &
  ~~~~~ ({\rm for}~ x>-r) \nonumber
\end{align}
}
We stress that Eqs.~(\ref{eq:J22(1)}),~(\ref{eq:J35}),~(\ref{eq:J36}) and~(\ref{eq:J37}) are valid
for all values of $x$ and $r$.
The constraints on the allowed values of $x$ and $r$ in Eqs. (\ref{eq:J41}) and (\ref{eq:J42}) are valid for our
calculation because $m_1$, the messenger fermion mass, is much larger than the gaugino mass.
We check that Eq.~(\ref{eq:J22(1)}) gives a consistent result of ref.~\cite{Davydychev:2000na}.
The functions $J_{35}(x,r)$ and $J_{37}(x,r)$ are compatible with the result of ref.~\cite{Czyz:2002re}
which uses differential equation method.

The higher transcendental functions with polylogarithm  up to third order are given by,
{\allowdisplaybreaks
\begin{align}
 J_{35}^{(1)}(x,r) &=
 \frac{5}{16} (12 r+23 x+12)
 +\frac{\pi^2}{24} (6 r+x+6+4\ln(r))
 -\frac{2\zeta (3)}{3} \left( \frac{r(9+5x)}{x}+5\right)
 \nonumber \\ &
 +\ln^2(r) \bigg(\,\frac{24 r x+r (r+12)+3 x^2-3}{8 x}
     +\frac{3}{4} (r-1) \ln (-x)\bigg)
 -\frac{1}{3}\ln^3(r)
 \nonumber \\ &
 +\Bigg[
 -\ln \left(\eta_1\,x+1\right) \Bigg(
   \frac{13}{4} \eta_2(r-x+1)
   +6 r
   +\frac{\pi^2}{6}(r+1)
   +\frac{13}{4} (x-1)
 \nonumber \\ & ~~~
 +\frac{3}{4} \left(\eta_1-\eta_2\right) (r-x+1)
  \ln \left(-\left(\eta_1-\eta_2\right){}^2 x^2\right) \Bigg)
 \nonumber \\ & ~~~
 +\ln^2\left(\eta_1\,x+1\right) \Bigg(
    \frac{1}{2} \eta _2 (-r+x-1)
    -\frac{3}{2} (r-1) \ln (-x)
  \nonumber \\ & ~~~ ~
  +\frac{r^2-6 r (x+1)-2 (x-1)x}{4 x}
  +\bigg(1-\frac{r (7 x+5)}{4 x}\,\bigg) \ln (r) \Bigg)
  \nonumber \\ & ~~~
  +\ln^3\left(\eta _1 x+1\right)\Bigg( \frac{r(3x+1)}{2x}-\frac{4}{3} \Bigg)
  +\text{Li}_2\left(\frac{1}{\eta _1}\right) \Bigg(
  \frac{r^2-1}{2 x}-3 r-x+5
   \nonumber \\ & ~~~ ~
  -2 \eta_2 (r-x+1)
  +\bigg(\frac{r(2x-1)}{x}-3\bigg) \ln (r)
  +\bigg(2-\frac{2r(2x+1)}{x}\bigg) \ln \left(\eta_2\,x+1\right)\Bigg)
   \nonumber \\ & ~~~
  +3(r-1)\text{Li}_2\bigg(\frac{\eta_2}{\eta_1}\bigg)
  \Bigg(\ln \left(\eta _2 \,x+1\right)- \frac{\ln(r)}{2}\Bigg)
   \nonumber \\ & ~~~
  +\text{Li}_2\left(\frac{\eta_2 \,x+1}{\eta_1\,x+1}\right) \Bigg(
      \frac{3}{4} (\eta_1-\eta_2) (r-x+1)
     +\frac{3r(x+1)}{x}\bigg(\frac{\ln(r)}{2} - \ln(\eta_2\,x+1) \bigg)
   \Bigg)
   \nonumber \\ & ~~~ ~
  +\text{Li}_3\bigg(\frac{1}{\eta_1}\bigg)
  \Bigg( \frac{2r(2x+3)}{x}+2\Bigg)
  +\text{Li}_3\bigg(\frac{1}{1-\eta _1}\bigg) \Bigg(\frac{2 r
   (x+3)}{x}+4\Bigg)
   \nonumber \\ & ~~~
  +\Bigg( \text{Li}_3\bigg(\,\frac{1-\eta _2}{1-\eta _1}\bigg)
  -       \text{Li}_3\bigg(\frac{\eta_2}{\eta _1}\bigg) \Bigg) \,\frac{3}{2}(r-1)
   \nonumber \\ & ~~~
  +\frac{3}{2} \text{Li}_3\bigg(\frac{\eta _2 x+1}{\eta _1 x+1}\bigg)
     \left(\frac{2 r}{x}+r+1\right)
  + (\eta_1 \leftrightarrow \eta_2 )
 \Bigg]
  \,,
  \\
 J_{37}^{(1)}(x,r) &=
 -\frac{11}{2}
 +\frac{\pi ^2}{12} \Big(2 \ln(r) +1 \Big)
 - \zeta (3) \bigg( \frac{6 r}{x}+ \frac{10}{3}\bigg)
 \nonumber \\ &
 - \ln^2(r) \bigg(\frac{3-5 r+3 x}{4 x}
    +\frac{3}{4}\ln(-x)\bigg)
 -\frac{1}{3} \ln^3(r)
 \nonumber \\ &
 +\Bigg[
 \ln \left(\eta_1\, x+1\right) \Bigg(
     5(1- \eta_2)
     -\frac{\pi^2}{6}
     -\frac{3}{2} ( \eta_1-\eta_2)
       \ln \left(-\left(\eta_1-\eta_2\right){}^2 x^2\right)\Bigg)
  \nonumber \\ &  ~~~
 +\ln^2\left(\eta_1\, x+1\right) \Bigg(
    1
    -\eta_2
    -\frac{r}{2 x}
    +\bigg(1-\frac{5 r}{4 x}\bigg) \ln(r)
    +\frac{3 }{2} \ln (-x)
    \Bigg)
 \nonumber \\ &  ~~~
 +\ln^3\left(\eta _1 \,x+1\right)\Bigg(\frac{r}{2 x}-\frac{4}{3}\Bigg)
 +\frac{3}{2}\, \text{Li}_2\left(\frac{\eta_2}{\eta_1}\right) \Big(
  \ln (r)-2 \ln \left(\eta_2\,x+1\right) \Big)
 \nonumber \\ &  ~~~
 +\text{Li}_2\left(\frac{1}{\eta_1}\right) \Bigg(
    4(1- \eta_2)
    +\frac{r-1}{x}
    -\left(\frac{r}{x}+3\right) \ln(r)
    +\bigg(2-\frac{2r}{x}\bigg) \ln \left(\eta_2\, x+1\right)
    \Bigg)
 \nonumber \\ &  ~~~
 +\frac{3}{2}\, \text{Li}_2\left(\frac{\eta _2 \,x+1}{\eta _1 \, x+1}\right)
  \Bigg(
     \eta_1-\eta_2
     +\frac{r}{ x} \bigg( \ln(r)-2 \ln \left(\eta _2 \,x+1\right) \bigg)
  \Bigg)
  \nonumber \\ &  ~~~
  +\text{Li}_3\bigg(\frac{1}{\eta_1}\bigg) \bigg(\frac{6r}{x}+2\bigg)
  +\text{Li}_3\bigg(\frac{1}{1-\eta_1}\bigg) \bigg(\frac{6 r}{x}+4\bigg)
  +\frac{3}{2} \text{Li}_3\bigg(\frac{\eta _2}{\eta _1}\bigg)
  \nonumber \\ &  ~~~
  -\frac{3}{2} \text{Li}_3\left(\frac{1-\eta _2}{1-\eta _1}\right)
  + \text{Li}_3\left(\frac{\eta_2 \,x+1}{\eta_1\,x+1}\right)
     \bigg(\frac{3 r}{x}+\frac{3}{2}\bigg)
  + (\eta_1 \leftrightarrow \eta_2 ) \Bigg] \,.
 \end{align}
 }

As for $J_{5i}$ $(i=1,2,3,4)$ we need transcendental functions higher
than polylogarithm of order 3. Their analytic expressions are beyond the scope of this paper.
Instead they can be given by a definite integral so that numerical evaluation of them is easily
performed.
To this end, we use the differential equation method in ref.~\cite{Kotikov:1990kg}.
For example, differentiating $I_{51}$ by $m_2^2$ is represented by a linear combination of other
master integrals in a diagrammatic way.
\begin{figure}[h]
\centerline{\epsfig{figure=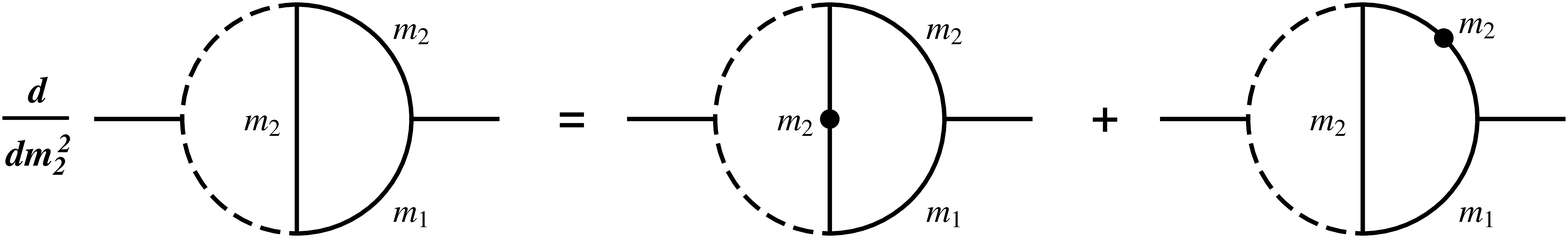, scale=0.2}}
\label{fig-MI-DE}
\end{figure}
By solving the differential equation, one can express the function $J_{51}$
as a definite integral.

After all, the functions $J_{5i}(i=1,2,3,4)$ are given by definite integrals of functions as follows,
\begin{eqnarray}
  J_{5i}(x,r) &=& \int_1^r dr' f_i(x,r') \left[ \sum_{j\geq1} c_{ij}(x,r') {\cal J}_{ij}(x,r')
  + c_{i0}(x,r')\right]  -\frac{{\hat J}_{i}(x)}{x}\,, \\
  J_{54}(x,r,\r2) &=& \int_1^r dr' f_4(x,r',\r2) \left[ \sum_{j\geq1} c_{4j}(x,r',\r2) {\cal J}_{4j}(x,r',\r2)
  + c_{40}(x,r',\r2)\right]
  \nonumber \\ &&
  +\int_1^\r2 dr' {\tilde f}_4(x,r') \left[ \sum_{j\geq1} {\tilde c}_{4j}(x,r') {\tilde {\cal J}}_{4j}(x,r') + {\tilde c}_{40}(x,r')\right]
   -\frac{{\hat J}_{4}(x)}{x},
\end{eqnarray}
where ${\cal J}_{ij}$ are a set of known master integral functions,
and ${\hat J}_{i}(x)$'s are the master integral functions in the limit that all the masses are equal.
One can find the expressions for ${\hat J}_{i}(x)$ in ref.~\cite{Broadhurst:1993mw,Fleischer:1998nb}:
 \begin{eqnarray}
 {\hat J}_1(x) &=&
  2 \ln^2(y) \ln (1-y)
 +6 \zeta(3)
 -6 \text{Li}_3(y)
 +6 \ln(y) \text{Li}_2(y) \,,\\
 {\hat J}_{2,3,4} (x) &=&
 6 \zeta (3)
 +12 \text{Li}_3(y)
 +24 \text{Li}_3(-y)
  -8\ln (y) (\text{Li}_2(y)+2 \text{Li}_2(-y))
 \nonumber \\ &&
  -2 \ln^2(y)(\ln(1-y)+2 \ln (1+y))\,.
 \end{eqnarray}
The coefficient functions $c_{ij}(x,r)$, the factor functions $f_i(x,r)$
and the corresponding set of master integrals ${\cal J}_{ij}(x,r)$ are given by,
 {\allowdisplaybreaks
 {\begin{align}
 f_1(x,r) &= r^{-1} x^{-3} (\eta_1-\eta_2)^{-2}\,, \nonumber \\
 {\cal J}_{1j}(x,r) &= \bigg\{
 J_{22}^{(1)}(x,r), J_{35}(x,r), J_{37}(x,r),
 J_{37}\bigg(\frac{x}{r},\frac{1}{r} \bigg),
 J_{41}\bigg(\frac{x}{r},\frac{1}{r} \bigg) \bigg\}\,, \nonumber \\
 c_{11}(x,r) &= - c_{15}(x,r)= \frac{x(x+4)}{r-1} + 3x\,, \nonumber \\
 c_{12}(x,r) &= -2(r+x+1)\,, \nonumber \\
 c_{13}(x,r) &= \frac{x(x+4)}{r-1} +4x +r +1\,, \nonumber \\
 c_{14}(x,r) &= -\frac{r x(x+4)}{r-1} +r(r+1)\,, \nonumber \\
 c_{10}(x,r) &=
 (1-r) (r+x+1) \text{Li}_2(1-r)
 +\frac{\pi^2}{12} (r+x+1)^2
 +\frac{11 x (x+4)}{2 (r-1)}
 \nonumber \\ &
 +\frac{1}{4} \bigg(37 r x+26 (r+1)^2+11 x^2+103 x\bigg)
 \nonumber \\ &
 +\ln(r) \Bigg(\,\frac{(1+2\eta_1)x(x+4)}{r-1}
       +\ln (x) \bigg(\eta_1\,x (r+x+1) +\frac{x(x+4)}{r-1}+3 x\bigg)
  \nonumber \\ & ~
        -(7 r-2) x - 6 r (r+1)-x^2+6 \eta_1 x
        \Bigg)
  \nonumber \\ &
 +\ln^2(r) \bigg(\,
   \frac{(\eta_1-1) x (x+4)}{r-1}-\eta_1 x (r+x-2)+\frac{3}{2} r(r+x+1)-3x\bigg)
 \nonumber \\ &
 +\frac{(1+y')}{(1-y')} \ln (y')
    \bigg(\,\frac{x(x+4)}{r-1}+3 x\bigg)
 +\ln^2(y') \bigg(\, \frac{(x+2) (x+4)}{2 (r-1)}
 \nonumber \\ & ~
   + 2 r^2+\frac{5 r(x+2)}{2}+\frac{(x+2)(x+4)}{2}\bigg)
 \nonumber \\ &
 -\left(\eta_1-\eta_2\right)\ln \left(\eta_1 x+1\right)
 \Bigg(\,\frac{2 x(x+4) }{r-1}+x (r+x+1) \ln (x) +6x
 \nonumber \\ & ~
 + \ln(r)\bigg(\,\frac{x (x+4)}{r-1}-x (r+x-2)\bigg)\Bigg)\,,
 \\
 f_2(x,r) &= r^{-1} x^{-4}(\eta_1-\eta_2)^{-4}\,, \nonumber \\
 {\cal J}_{2j}(x,r) =& \bigg\{
 J_{22}^{(1)}(x,r),
 J_{35}(x,r),
 J_{37}(x,r),
 J_{37}\bigg(\frac{x}{r},\frac{1}{r} \bigg),
 J_{35}^{(1)}(x,r),
 J_{37}^{(1)}(x,r),
 J_{37}^{(1)}\bigg(\frac{x}{r},\frac{1}{r} \bigg)
 \bigg\}\,, \nonumber \\
 c_{21}(x,r)&=(\eta_1-\eta_2)^2 x^2(r-x-1)\,, \nonumber \\
 c_{22}(x,r)&= 15r^2+8r(x-1)-7(x+1)^2 \,,\nonumber \\
 c_{23}(x,r)&= -4(r^2+4rx-(x+1)^2)\,,\nonumber \\
 c_{24}(x,r)&= (-r+x+1)\Big( 9r^2+2r(5x-1)+(x+1)^2\Big) - 4r(r-x-1)(r+x)\ln(r)\,, \nonumber \\
 c_{25}(x,r)&= 2(-r+x+1)(r+x+1) \,,\nonumber \\
 c_{26}(x,r)&= 2(x+1)(r-x-1) \,,\nonumber \\
 c_{27}(x,r)&= 2r(r-x-1)(r+x) \,,\nonumber \\
 c_{20}(x,r)&=
 \frac{4}{3} \zeta (3) r(-r+x+1)
 -\frac{\pi^2}{12} \Big(3 r^3+r^2 (7 x+3)+r (x+1) (5x-7)+(x+1)^3\Big)
 \nonumber \\ &
 -35 r^3-r^2\bigg(\frac{77}{2} x+9\bigg)- r (x-5) (2 x+7) +\frac{3}{2} (x+1)^2 (x+6)
 \nonumber \\ &
 +\ln(r)\bigg(33 r^3-r^2 (25-27 x) -\frac{1}{3} \pi^2 r (r-x-1)
 \nonumber \\ & ~
 -r (x+1) (5 x+9)+(x+1)^3\bigg)
 +\ln ^2(r) \bigg(\eta _1 \left(\eta_1-\eta _2\right){}^2 x^2(r+x+1)
 \nonumber \\ & ~
 -\frac{1}{2} \Big (19 r^3+3 r^2 (5 x-7)-r (x+1) (3 x-1)+(x+1)^3\Big)\bigg)
 \nonumber \\ &
 -\frac{1}{3} \ln ^3(r) r (r-x-1) (4 r+4 x+1)
 \nonumber \\ &
 +\left(\eta _1-\eta _2\right){}^2 \ln \left(\eta_1\, x+1\right)  \ln (r)
  \bigg(x(-r+x+1)(r+x-1) -2 \eta_1 x^2(r+x+1)\bigg)
 \nonumber \\ &
 +2\left(\eta _1-\eta _2\right){}^2 \ln^2\left(\eta _1 \,x+1\right) r x(r+x-1)\,,
  \\
 f_3(x,r) &= (2r+x+2)^{-1} x^{-2}(\eta_1-\eta_2)^{-2}\,, \nonumber \\
 {\cal J}_{3j}(x,r) &= \bigg\{
 J_{22}^{(1)}(x,r),
 J_{41}(x,r),
 J_{41}\bigg(\frac{x}{r},\frac{1}{r} \bigg)
 \bigg\}\,, \nonumber \\
 c_{31}(x,r)&= -\frac{2x(x+4)}{r-1} -2x \,, \nonumber \\
 c_{32}(x,r)&= \frac{2x(x+4)}{r-1}+2x-4 \,,\nonumber \\
 c_{33}(x,r)&= 4 \,,\nonumber \\
 c_{30}(x,r)&=
 -4 \text{Li}_2(1-r) (r+x+3)
 -\frac{11 x(r+x+3)}{r-1}
 -\ln (r) \left(\frac{4 \left(\eta_1-1\right)x(r+x+3)}{r-1}+16\right)
   \nonumber \\ & ~
 -\ln ^2(r) \left(2 \eta_1(r+x+5)+\frac{(x+4)^2}{r-1}+2 r+3 x+8\right)
  \nonumber \\ & ~
 +\ln \left(\eta _1 \,x+1\right) \Bigg(\ln (r)
     \bigg(\frac{2 \left(r(x-2)+x^2+x+2\right)}{x}
     +4 \eta _1 (r+x+5)
  \nonumber \\ & ~
     +\frac{2 (x+4) (r+x+3)}{r-1}\bigg)
     +\frac{4 \left(\eta _1-\eta _2\right) x (r+x+3)}{r-1}\Bigg)
 \nonumber \\ & ~
 -2 \ln ^2\left(\eta _1 x+1\right) \frac{  \left(\eta _1-\eta _2\right){}^2 x
   (r+x+3)}{r-1}
 -2\frac{(1+y)}{1-y}\ln (y)\left(\frac{x(x+4)}{r-1}+x-2\right)
  \nonumber \\ & ~
 -4\frac{ (1+y')}{1-y'} \ln (y')
  +\ln^2(y)\left(\frac{2 (x+4)}{r-1}+\frac{4}{x}+4\right)
 \nonumber \\ & ~
  +\ln^2(y')\left(\frac{4 r (x-1)}{x}+\frac{(x+4) (2 r+x+2)}{r-1}+3 (x+2)\right) \,,
 \\
 f_4(x,r,\r2) &= \Big(x(r-1)(\r2-1)+(r+\r2-2)(r\r2-1)\Big)^{-1}
              x^{-2}(\eta_1-\eta_2)^{-2}\,, \nonumber \\
 {\cal J}_{4j}(x,r,\r2) &= \bigg\{
 J_{22}^{(1)}(x,r),
 J_{22}^{(1)}(x,\r2),
 J_{35}\bigg(\frac{x}{r},\frac{\r2}{r} \bigg),
 J_{37}\bigg(\frac{x}{r},\frac{\r2}{r} \bigg),
 J_{37}\bigg(\frac{x}{\r2},\frac{r}{\r2} \bigg),
 J_{41}(x,r),
 J_{41}(x,\r2),
  \nonumber \\ & ~~~
 J_{42}(x,r,\r2),
 J_{42}(x,\r2,r)
 \bigg\}\,, \nonumber \\
 c_{41}(x,r,\r2)&= r^2\r2 + \frac{x(x+4)(\r2-1)}{r-1}+r(x\r2-\r2-x-1)
                +x(3\r2+x+1) +1 \,, \nonumber \\
 c_{42}(x,r,\r2)&= -x^2(\eta_1-\eta_2)^2
                \bigg( \frac{\r2-1}{r-1}+\r2\bigg)  \,,\nonumber \\
 c_{43}(x,r,\r2)&= 2r\Big( x(1-\r2)-r\r2-r-\r2+3\Big) \,,\nonumber \\
 c_{44}(x,r,\r2)&= \frac{x(x+4)(\r2-1)}{r-1}
                +x(2r\r2-3r+2\r2)
                +r(2r\r2 +r +\r2-5)
                -\r2-x+2 \,,\nonumber \\
 c_{45}(x,r,\r2)&= \r2 \Big( r(\r2-2x+3)+(x+1)\r2-r^2-x(x+3)-4 \Big) \,,\nonumber \\
 c_{46}(x,r,\r2)&= -\frac{x(x+4)(\r2-1)}{r-1}
                -x(r\r2-2r+2\r2-1)
                -(r-1)(r\r2+\r2-2)
                \,,\nonumber \\
 c_{47}(x,r,\r2)&= x^2(\eta_1-\eta_2)^2\r2 \,,\nonumber \\
 c_{48}(x,r,\r2)&= -x^2-x(r+\r2+2)+r\r2-r-\r2+1 \,,\nonumber \\
 c_{49}(x,r,\r2)&= x^2(\eta_1-\eta_2)^2\,\frac{\r2-1}{r-1} \,,\nonumber \\
 c_{40}(x,r,\r2)&=
   (r-1)\text{Li}_2(1-r) \left(3+x-r - \tilde{r} (r+x+1)\right)
   \nonumber \\ &
   +\text{Li}_2\left(1-\r2\right) \left(r-x-3 -\r2^2 (r+x+1)+2 \r2(x+2)\right)
   \nonumber \\ &
   +\frac{1}{4} \left(26 r^2-r (39 x+82)+x (9 x+25)+4\right)
   +\frac{ x (x+4)(\r2-1)}{2(r-1)}
   \nonumber \\ &
   +\frac{\tilde{r}}{4}  \left(26 r^2+r (15 x+56)-x (11 x+35)-82\right)
   +\frac{13}{2} \tilde{r}^2 (r+x+1)
   \nonumber \\ &
   +\frac{\pi^2}{12}
   \bigg(\tilde{r}^2 (r+x+1)+\tilde{r} \left((r+x)^2-x-1\right)
   -\frac{x (x+4)(\r2-1)}{r-1}
   \nonumber \\ & ~
   +(r+1) (r+x-2)\bigg)
   +\ln (r) \bigg(\tilde{r} \left(r^2+2 r (x+3)+(x-3) x-19\right)
   \nonumber \\ & ~
       +6 \tilde{r}^2 (r+x+1)
       +\frac{x (x+4)(\r2-1)}{r-1}
       -r (x+2)-x (x+4)+2\bigg)
   \nonumber \\ &
   - \ln \left(\tilde{r}\right) \Big(
     \tilde{r} \left(r^2+2 r (x+2)+(x-4) x-17\right)
     +6 \tilde{r}^2 (r+x+1)\Big)
   \nonumber \\ &
   +\ln^2(r) \bigg(2 \tilde{r}^2 (r+x+1)
       +\frac{1}{2} \tilde{r} (3 r-2 x-13)
       +\frac{x (x+4)(\r2-1)}{2(r-1)}
   \nonumber \\ & ~
       -\frac{1}{2} (r+1) (r+x-2)\bigg)
   + \ln^2\left(\tilde{r}\right)
       \bigg(\frac{1}{2} \tilde{r} \left(-r^2+r (5-2 x)-x (x+5)-10\right)
   \nonumber \\ & ~
       +\frac{3}{2} \tilde{r}^2 (r+x+1)\bigg)
   +4\ln (r)\ln \left(\tilde{r}\right)
   \left( \tilde{r} (-r+x+3)-\tilde{r}^2 (r+x+1)\right)
  \nonumber \\ &
 +\frac{1+y}{1-y}\ln y
  \Bigg(\tilde{r} \left(-r (x-2)-x^2-2\right)
    +\frac{x (x+4)(\r2-1)}{r-1}
    -2 r (x+1)-x+2\Bigg)
   \nonumber \\ &
    +\ln^2 y
  \Bigg(\tilde{r} \bigg(\frac{2 (r-1)}{x}+2 r+\frac{3x}{2}+1\bigg)
  +\frac{(x+2)(x+4)(\r2-1)}{2(r-1)}
 \nonumber \\ & ~
  -\frac{r}{2x} (x^2+2x+4)
  -\frac{ (x+2) (x+6)}{2}
  +\frac{2}{x} \Bigg)
  \nonumber \\ &
  +\Big((\eta_1-1)\ln(\eta_2\,x+1)+(\eta_2-1)\ln(\eta_1\,x+1)\Big)
   \Big((\tilde{\eta}_1-1)\ln(\tilde{\eta}_2\,x+1)
  \nonumber \\ & ~
      +(\tilde{\eta}_2-1)\ln(\tilde{\eta}_1\,x+1) \Big)
      \Big( x(-r+x+3)- x \tilde{r} (r+x+1)\Big)
   \nonumber \\ &
 +\Big((\eta_1-1)\ln(\eta_2\,x+1)+(\eta_2-1)\ln(\eta_1\,x+1)\Big)
   \Bigg(
     -\left(\eta _1-\eta _2\right){}^2 x^2 \tilde{r} \ln \left(\r2\right)
   \nonumber \\ & ~
     +\frac{2 x (x+4)(\r2-1)}{r-1}
     +\tilde{r} (2 r(r+x-1)+6 x)
     +2 \left(-r (x+1)+x^2+x+1\right)
     \Bigg)
   \nonumber \\ &
 +\Big((\tilde{\eta}_1-1)\ln(\tilde{\eta}_2\,x+1)
      +(\tilde{\eta}_2-1)\ln(\tilde{\eta}_1\,x+1) \Big)
  \nonumber \\ & ~
  \times \Bigg(\ln (r)
   \bigg(\tilde{r} \left(r^2+(r+2) x-1\right)
     +\frac{x (x+4)(\r2-1)}{r-1}
     -2 r(x+1)-x+2\bigg)
  \nonumber \\ & ~
   -2 x^2\left(\eta _1-\eta _2\right){}^2 \
   \bigg(\frac{\r2-1}{r-1}+ \tilde{r}\bigg)\Bigg)\,,
 \\
 \tilde{f}_4(x,r) &= (r-1)^{-1} x^{-2}(\eta_1-\eta_2)^{-2}\,, \nonumber \\
 {\tilde{ \cal {J}}}_{4j}(x,r) &= \bigg\{
 J_{22}^{(1)}(x,r),
 J_{35}(x,r),
 J_{37}(x,r),
 J_{37}\bigg(\frac{x}{r},\frac{1}{r} \bigg),
 J_{41}(x,r)
 \bigg\}\,, \nonumber \\
 \tilde{c}_{41}(x,r)&=- \tilde{c}_{45}(x,r)= r-x-1\,, \nonumber \\
 \tilde{c}_{42}(x,r)&= -4\,, \nonumber \\
 \tilde{c}_{43}(x,r)&= -r-x+3\,, \nonumber \\
 \tilde{c}_{44}(x,r)&= 3r-x-1\,, \nonumber \\
 \tilde{c}_{40}(x,r)&=
 2(1- r)\text{Li}_2(1-r)
 +\frac{\pi^2}{6} (r+x+1)
 +\frac{37r}{2} + \frac{15}{2}
 -\ln(r) (11 r+x+1)
 \nonumber \\ &
 +\frac{1}{2} \ln^2(r) (7 r-x-1)
 +\frac{1+y}{1-y} \ln (y)(r-x-1) (\ln (r)+1)
 \nonumber \\ &
 +\ln^2(y)\bigg(\frac{r-1}{x}+ \frac{r+x+5}{2} \bigg)
 +\Big((\eta_1-1)\ln(\eta_2\,x+1)+(\eta_2-1)\ln(\eta_1\,x+1)\Big)
 \nonumber \\ & ~
 \times \left(2 (r-x-1)-2 x\frac{(1+y)}{1-y} \ln (y)\right)\,,
 \end{align}
 }
 where the new variables are defined by
 \begin{eqnarray}
 y' \equiv y |_{x \to x/r}\,,
 ~~~ \tilde{\eta}_1 \equiv \eta_1 |_{r \to \r2}\,,
 ~~~ \tilde{\eta}_2 \equiv \eta_2 |_{r \to \r2}\,.
 \end{eqnarray}

For the calculation of the gluino pole mass,
setting $x\to 0$ leads to a good approximation of the self-energy function $\Omega^{(2)}$.
In that limit, we can do the above integrations so that
the analytic expressions of $J_{5i}$ are given as follows:
 \begin{eqnarray}
 J_{51}(0,r)&=&
 -\frac{(r+1)}{(r-1)^2}\Li_2(1-r)
 -\frac{4}{r-1}
 +\frac{(3-r)}{(r-1)^2}\ln(r)
 -\frac{\ln^2(r)}{2(r-1)}
 \nonumber \\ &&
 +\ln(x)\Bigg( \frac{1}{r-1}-\frac{\ln(r)}{(r-1)^2} \Bigg) \,,
 \\
 J_{52}(0,r) &=& J_{53}(0,r) =
 \frac{2}{r-1} \Li_2 (1-r)
 +\frac{r}{(r-1)^2} \ln^2(r) \,,
  \\
 \label{eq:J54x0}
 J_{54}(0,r,\r2) &=&
 \frac{\Li_2(1-r)}{\r2-1}
 +\frac{\Li_2(1-\r2)}{r-1}
 +\frac{\Li_2(\r2)}{\r2-1}
 \nonumber \\ &&
 -\frac{\pi^2}{6(r-1)}
 +\frac{r\ln(r)\ln(\r2)}{(r-1)(\r2-1)}
 +\frac{\ln(1-\r2)\ln(\r2)}{\r2-1}
 \nonumber \\ &&
 -\frac{(r-\r2)}{(r-1)(\r2-1)} \bigg(\,
     \Li_2\Big(\frac{\r2}{r}\Big)
     + \ln\Big(\frac{\r2}{r}\Big) \ln(r-\r2)
     +\frac{\ln^2(r)}{2}
 \bigg)\,.
 \end{eqnarray}
It should be noted that there remains a singularity of $\ln(x)$ in $J_{51}(x,r)$ as $x\to0$.
We comment that the singularity can be handled using
the asymptotic expansion technique~\cite{Davydychev:1992mt,Berends:1994sa}.

\section{Explicit formula for $\Omega^{(2)}(s)$\label{app:fullomega}}
Using the master integrals in Appendix~\ref{app:masterint},
we can write down the full expression of the self-energy function $\Omega^{(2)}$
as follows,
{\allowdisplaybreaks
{\begin{align}
\Omega^{(2)}(s) &=
\bigg(\frac{\alpha_s}{4\pi}\bigg)^2 \Lambda(r-1)^{-1} N_{\rm{mess}} C(R) C_2(R)
 \Bigg[
  \nonumber \\ & ~~
   -4J_{22}^{(1)}(x,r)\bigg(
     \frac{ r- x+1}{r}
    +\frac{ \left((x+1)^2+r (x-5)\right)}{\left(\eta _1-\eta _2\right){}^2 r x}
      \bigg)
 \nonumber \\ & ~~
   +4J_{35}\bigg(\frac{x}{r},\frac{\tilde{r}}{r}\bigg)
     \bigg(
      r-x+2
    +\frac{ (x-2) (x+1)^2+r \left(x^2-9 x+2\right)}
          {\left(\eta _1-\eta _2\right){}^2 x^2}\bigg)
 \nonumber \\ & ~~
   +8 J_{36}(x,r)
   +4 J_{37}(x,r)
   +4 J_{37}\bigg(\frac{x}{r},\frac{1}{r}\bigg)
  \nonumber \\ & ~~
  - 2 J_{37}\left(\frac{x}{r},\frac{\tilde{r}}{r}\right)
  \bigg(
    2r-3x+4
    +\frac{ 2\left((x-1) (x+1)^2+r \left(x^2-7 x+1\right)\right)}
       {\left(\eta_1-\eta _2\right){}^2 x^2}\bigg)
  \nonumber \\ & ~~
   +2J_{37}\left(\frac{x}{\tilde{r}},\frac{r}{\tilde{r}}\right)
      \frac{\left(x^3+2 x^2+3 x+r \left(x^2-7 x-2\right)+2\right)}
      {\left(\eta _1-\eta_2\right){}^2 x^2}
  \nonumber \\ & ~~
  +4 J_{41}(x,r) \bigg(
      \frac{ r-x+1}{r}
      +\frac{ (x+1)^2+r (x-5)}{\left(\eta _1-\eta _2\right){}^2 r x}
      \bigg)
  \nonumber \\ & ~~
  +2J_{41}\left(\frac{x}{r},\frac{1}{r}\right)
   \bigg( r+3-\frac{ 3 r+x+1}{\left(\eta _1-\eta_2\right){}^2 x}
   \bigg)
  +2 J_{42}\left(x,r,\tilde{r}\right)
  \bigg(1-r+\frac{ 3 r+x+1}{\left(\eta _1-\eta _2\right){}^2 x}\bigg)
  \nonumber \\ & ~~
 -4 (r+x+1) J_{52}(x,r)
 +2(r-1)(J_{54}(x,r,\r2)-J_{53}(x,r) )
  \nonumber \\ & ~~
 +\frac{2\text{Li}_2(1-r)}{r}\bigg(
 2(r+4)x+3
 +\frac{x^3+(r-4) x^2-(17 r+13) x-8}
    {\left(\eta _1-\eta _2\right){}^2  x}
 -\left(\eta _1-\eta _2\right){}^2 x^2
 \bigg)
 \nonumber \\ & ~~
 -2\text{Li}_2\left(1-\tilde{r}\right)
  \bigg(
  r
  +\frac{x^2+(r+1)x-8r}{\left(\eta _1-\eta _2\right){}^2x}\bigg)
 +\frac{\pi ^2}{6} \bigg( 2r+\frac{(x+1) \left(x^2+3 x+r
   (x-4)+4\right)}{ \left(\eta _1-\eta _2\right){}^2 x^2}\bigg)
 \nonumber \\ & ~~
 -\frac{2 \left(\eta _1-\eta _2\right){}^2 x^2}{r}
 +\frac{15 x^2+88 x-144}{2 r}
  \nonumber \\ & ~~
 -\frac{\left(11 x^2+80 x-104\right) (x+1)^2+r \left(11
   x^3+7 x^2-680 x+104\right)}
 {2 \left(\eta _1-\eta _2\right){}^2 r x^2}
 \nonumber \\ & ~~
 +2\ln(r)\bigg(
   2r-12x-19
   +\frac{r \left(6 x^2-23 x-12\right)+3 \left(2 x^3+7 x^2+9 x+4\right)}
   {\left(\eta _1-\eta _2\right){}^2 x^2}
     \bigg)
 \nonumber \\ & ~~
 -12\ln \left(\tilde{r}\right)\bigg(
  \frac{7r}{6}
  +\frac{x^3+2 x^2+3 x+r \left(x^2-7 x-2\right)+2}
   {\left(\eta_1-\eta _2\right){}^2 x^2}\bigg)
\nonumber \\ & ~~
  -2 \ln (r) \ln\left(\tilde{r}\right)
  \bigg(
    4r
    +\frac{4 x^3+7 x^2+12 x+r \left(4 x^2-31 x-9\right)+9}
    {\left(\eta_1-\eta _2\right){}^2 x^2}\bigg)
 \nonumber \\ & ~~
  +\ln ^2(r)\bigg(
    2(r+1)
   +\frac{3 x \left(x^2-1\right)+r \left(3 x^2-27 x+16\right)}
   {\left(\eta _1-\eta _2\right){}^2 x^2}\bigg)
 \nonumber \\ & ~~
 + \ln ^2\left(\tilde{r}\right)\bigg(
  2 r
  +\frac{3 \left(x^3+2 x^2+3 x+r \left(x^2-7 x-2\right)+2\right)}
  {\left(\eta _1-\eta_2\right){}^2 x^2}\bigg)
 \nonumber \\ & ~~
 +\frac{4(1+y)}{1-y}\ln (y)
   \bigg(\frac{ x-2}{r}
   -\frac{ (x+1)^2+r (x-5)}{\left(\eta _1-\eta_2\right){}^2 r x}
   \bigg)
 \nonumber \\ & ~~
 +\frac{1+y'}{1-y'}\ln (y') \left(\frac{2 (3 r+x+1)}{\left(\eta _1-\eta_2\right){}^2
   x}-4 r\right)
 \nonumber \\ & ~~
 +\ln^2(y)
   \bigg(\frac{2 \left(3 x^2+5 r x+9 x+2 r+6\right)}
    {\left(\eta _1-\eta _2\right){}^2 rx^2}
    -\frac{8 (x+1)}{r x}\bigg)
 +\ln^2(y') \Bigg(
 x+\frac{4}{x}+22
   \nonumber \\ & ~~ ~
   +r \bigg(10-\frac{8}{x}\bigg)
   +\frac{(3 x-14) (x+1)^2+r \left(3 x^2-41 x+22\right)}
       {x^2 \left(\eta _1-\eta _2\right){}^2}
  -4 x \left(\eta _1-\eta _2\right){}^2
       \Bigg)
 \nonumber \\ & ~~
 +4\Big((\eta_1-1)\ln(\eta_2\,x+1)+(\eta_2-1)\ln(\eta_1\,x+1)\Big)^2
  \nonumber \\ & ~~
 -\Big((\eta_1-1)\ln(\eta_2\,x+1)+(\eta_2-1)\ln(\eta_1\,x+1)\Big)
 \Bigg(
   \frac{4 (11 r-4 x+1)}{r}
   \nonumber \\ & ~~ ~
   +\frac{16 \left((x+1)^2+r(x-3)\right)}{\left(\eta _1-\eta _2\right){}^2 r x}
   +\ln (r)\bigg(2 r+\frac{4 \left(x^2+3x-2 r+2\right)}
      {\left(\eta _1-\eta _2\right){}^2 x^2}\bigg)
   \nonumber \\ & ~~ ~
   -\ln \left(\tilde{r}\right)
   \bigg(\frac{4(r-2)}{\left(\eta _1-\eta _2\right){}^2 x}
   -2 r+4\bigg) \Bigg)
    + 4\ln \left(\frac{\mu ^2}{M^2_{\rm mess}}\right)
  \Bigg(
  \ln(r)\bigg(\frac{r+x+1}{\left(\eta _1-\eta _2\right){}^2x}-1\bigg)
   \nonumber \\ & ~~~
 +\Big((\eta_1-1)\ln(\eta_2\,x+1)+(\eta_2-1)\ln(\eta_1\,x+1)\Big)
    \bigg(\frac{r+x-1}{\left(\eta _1-\eta_2\right){}^2 x}-2\bigg)
  \Bigg)
  \nonumber \\ & ~~
   - (r \leftrightarrow \r2 )
   \Bigg]
\nonumber \\ &
+\bigg(\frac{\alpha_s}{4\pi}\bigg)^2 \Lambda(r-1)^{-1} N_{\rm{mess}} C(R) C_2(G)
 \Bigg[
 \nonumber \\ & ~~
 -3J_{22}^{(1)}(x,r)
 -\frac{16}{x} J_{35}(x,r)
 +J_{37}(x,r)\bigg(2+\frac{8}{x}-\frac{8}{r-1}\bigg)
 \nonumber \\ & ~~
 +J_{37}\left(\frac{x}{r},\frac{1}{r}\right)
   \bigg(
     \frac{ 8x \left(\eta _1-\eta_2\right){}^2}{r-1}
    -\frac{ 8x^2+9rx+15x-8r+8}{(r-1) x}
   \bigg)
 \nonumber \\ & ~~
 +2J_{41}(x,r)
 +J_{42}(x,r,\r2)
 +J_{51}(x,r) (2 -2 r+x)
 +4 J_{51}\left(\frac{x}{r},\frac{1}{r}\right) \left(1-\frac{1}{r}\right)
  \nonumber \\ & ~~
 +2 J_{52}(x,r)  (r+x+1)
 +(r-1) (J_{53}(x,r)-J_{54}(x,r,\r2))
 +8\text{Li}_2(1-r)\Big(\frac{1-r}{x}\Big)
   \nonumber \\ & ~~
 +\frac{2 \pi ^2 r}{3 x}
 +\frac{52 r}{x}
 +\ln (r)
   \bigg(
   97
   -\frac{48}{x}
   +\frac{48x+200}{r-1}
   -\frac{48 \left(\eta _1-\eta _2\right){}^2 x}{r-1}
   -\ln (x) \Big(\frac{6r+2}{r-1}\Big)
  \bigg)
  \nonumber \\ & ~~
 +\frac{\ln^2(r)}{r-1} \bigg(
    12 x \left(\eta _1-\eta_2\right){}^2
    -\frac{24x^2+43rx+45x-24r+24}{2x}
    \bigg)
 \nonumber \\ & ~~
   +\frac{6(1+y')}{1-y'}\ln(y')
  +\ln^2(y') \bigg(1-\frac{2r}{x}\bigg)
  \nonumber \\ & ~~
  -2\Big((\eta_1-1)\ln(\eta_2\,x+1)+(\eta_2-1)\ln(\eta_1\,x+1)\Big)^2
  \nonumber \\ & ~~
  -\Big((\eta_1-1)\ln(\eta_2\,x+1)+(\eta_2-1)\ln(\eta_1\,x+1)\Big)
  \bigg( 22 + \ln(r) - 6\ln(x)
  \nonumber \\ & ~~ ~
  +8\ln\Big(\frac{\mu^2}{M^2_{\rm{mess}}}\Big)
  \bigg)
   - (r \leftrightarrow \r2 )
  \Bigg] .
 \end{align}
 }


\end{document}